\documentclass[preprint,12pt]{elsarticle}

\usepackage{listings}
\usepackage{color}

\definecolor{dkgreen}{rgb}{0,0.6,0}
\definecolor{gray}{rgb}{0.5,0.5,0.5}
\definecolor{mauve}{rgb}{0.58,0,0.82}

\usepackage{tikz}
\usetikzlibrary{backgrounds}
\usepackage{graphicx}
\usepackage{bm}
\usepackage{amsmath}
\usepackage{setspace}
\usepackage{hyperref}
\usepackage{subfigure}
\usepackage{multirow}
\hypersetup{
    colorlinks=true,       
}
\usepackage{amssymb}

\usepackage{lineno}
\usepackage{diagbox}

\biboptions{sort&compress}

\journal{Journal Name}

\begin{document}

\begin{frontmatter}


\title{Hidden Fluid Mechanics: A Navier-Stokes Informed Deep Learning Framework for Assimilating Flow Visualization Data}



\author{Maziar Raissi, Alireza Yazdani, and George Em Karniadakis}
\address{Division of Applied Mathematics, Brown University,\\ Providence, RI, 02912, USA}


\begin{abstract}
We present \emph{hidden fluid mechanics} (HFM), a physics informed deep learning framework capable of encoding an important class of physical laws governing fluid motions, namely the Navier-Stokes equations. In particular, we seek to leverage the underlying conservation laws (i.e., for mass, momentum, and energy) to infer hidden quantities of interest such as velocity and pressure fields merely from spatio-temporal visualizations of a passive scaler (e.g., dye or smoke), transported in arbitrarily complex domains (e.g., in human arteries or brain aneurysms). Our approach towards solving the aforementioned data assimilation problem is unique as we design an algorithm that is agnostic to the geometry or the initial and boundary conditions. This makes HFM highly flexible in choosing the spatio-temporal domain of interest for data acquisition as well as subsequent training and predictions. Consequently, the predictions made by HFM are among those cases where a pure machine learning strategy or a mere scientific computing approach simply cannot reproduce. The proposed algorithm achieves accurate predictions of the pressure and velocity fields in both two and three dimensional flows for several benchmark problems motivated by real-world applications. Our results demonstrate that this relatively simple methodology can be used in physical and biomedical problems to extract valuable quantitative information (e.g., lift and drag forces or wall shear stresses in arteries) for which direct measurements may not be possible.
\end{abstract}


\end{frontmatter}



\section{Introduction}

Recent advances in deep learning and computational resources in addition to new data recordings and sensor technologies have the potential to revolutionize our understanding of the physical world in modern application areas such as image recognition \cite{krizhevsky2012imagenet}, drug discovery \cite{gawehn2016deep}, and bioinformatics \cite{alipanahi2015predicting}. Moreover, many concepts from deep learning can be integrated with classical methods in mathematical physics to help us solve sophisticated data assimilation and inverse problems involving complex dynamical systems. This integration of nonlinear dynamics and deep learning opens the door for principled methods for model construction and predictive modeling. Furthermore, recent developments in \emph{physics-informed deep learning} have enabled us to train deep neural networks with millions of parameters using very few training examples, for cases where the available data are known to respect a given physical law described by a system of partial differential equations \cite{raissi2017physics_I,raissi2017physics_II}. The prior knowledge of an underlying physical law introduces important structure that effectively regularizes the minimization procedure in the training of neural networks, and enables them to generalize well even when only a few training examples are available. An important class of these physical laws govern the fluid motions in nature, which will be the focus of the present work.\\

Quantifying the flow dynamics essentially requires the knowledge of spatio-temporal fluid velocity and pressure fields, and has been the centerpiece of experimental and theoretical fluid mechanics for centuries. Traditionally, this has been achieved by measuring the instantaneous fluid velocity and pressure experimentally, or by solving the governing equations of fluid mechanics in a precisely defined geometry with proper initial and boundary conditions. Despite significant advances in experimental fluid mechanics (e.g., with the emergence of particle image velocimetry), the use of measurements to reliably infer fluid velocity and pressure/stress fields is not a straightforward task. Furthermore, although experimental measurements of external flows (e.g., flow past a bluff object) are obtained relatively easily, albeit in small subdomains, leading to two or even three dimensional vector fields for velocity, the quantification of velocity fields for internal flows (e.g., blood flow in the vascular networks) could become very difficult or impractical. From the theoretical standpoint, the governing equations of fluid mechanics have been derived from conservation laws (i.e., conservation of mass, momentum and energy) leading to partial differential equations such as the well-known Navier-Stokes equations for Newtonian fluids \cite{batchelor2000introduction}. With the increasing computational power, accurate solutions of such equations in a ``forward" setting are now available with lower cost even in the turbulence regime using Direct Numerical Simulations (DNS), hence resolving all energetically important scales down to Kolmogorev dissipation scales. However, the assimilation of experimental fluid mechanics data into the mathematical models to infer velocity and pressure fields in an ``inverse" setting has not been adequately addressed so far. The main objective of this article is to extend and incorporate the aforementioned physics-informed deep learning algorithms \cite{raissi2017physics_I,raissi2017physics_II} to leverage the hidden physics of fluid mechanics to infer the latent quantities of interest (e.g., the velocity and pressure fields) through minimalistic usage of data obtained by the ``passive scalar" transport (e.g., transport of dye or smoke).\\

Let us consider the transport of a passive scalar field $c$ by a velocity field $\bm{u}$. Such a problem arises, for example, while studying the spreading of smoke or dye advected by a given velocity field and subject to molecular diffusion. We expect the dynamics of $c$ to be governed by an effective transport equation \cite{vergassola1997scalar} written in the form of
\begin{equation}\label{eq:transC}
\rho \left(\partial_t c + \bm{u} \cdot \nabla c \right) = \kappa \nabla^2 c,
\end{equation}
where $\kappa$ is the molecular diffusivity and $\rho$ is the density. Transport of scalar fields in fluid flow has been studied in numerous applications such as aerodynamics, biofluid mechanics, and non-reactive flow mixing to name a few. The use of smoke in wind tunnels or dye in water tunnels for flow visualization and quantification has long been practiced in experimental fluid mechanics \cite{barlow2015low}. Moreover, recent techniques in planar laser induced fluorescence imaging combined with particle image velocimetry have been developed to assess the relationships between scalar and velocity/vorticity fields \cite{koochesfahani1985laser,crimaldi2008planar}. The use of scalar transport in conjunction with advanced imaging modalities to quantify blood flow in the vascular networks is now a common practice. For example, coronary computed tomography (CT) angiography is typically performed on multidetector CT systems after the injection of non-diffusible iodine contrast agent, which allows coronary artery visualization and the detection of coronary stenoses \cite{voros2011coronary}. Another example is the quantification of cerebral blood flow, which is detrimental in the prognostic assessments in stroke patients using a contrast agent and perfusion CT \cite{wintermark2002prognostic}, and in cognitive neuroscience with the use of functional magnetic resonance imaging that only relies on the blood-oxygen-level dependent contrast to measure brain activity \cite{aron2007triangulating}.\\

Inspired by recent developments in \emph{physics-informed deep learning} \cite{raissi2017physics_I,raissi2017physics_II} and \emph{deep hidden physics models} \cite{raissi2018deep}, we propose to leverage the hidden physics of fluid mechanics (i.e., the Navier-Stokes equations) and infer the latent quantities of interest (e.g., the velocity and pressure fields) by approximating them using deep neural networks. This choice is motivated by modern techniques for solving forward and inverse problems associated with partial differential equations, where the unknown solution is approximated either by a neural network \citep{raissi2017physics_I,raissi2017physics_II,raissi2018forwardbackward,raissi2018multistep} or a Gaussian process \citep{raissi2018numerical,raissi2018hidden,raissi2017inferring,raissi2017machine,raissi2017parametric,perdikaris2017nonlinear,raissi2016deep}. Moreover, placing a prior on the solution itself is fully justified by the similar approach pursued in the past century using classical methods of solving partial differential equations such as finite elements and spectral methods, where one would expand the unknown solution in terms of an appropriate set of basis functions. Our focus here is the transport of a passive scalar by incompressible Newtonian flows in both unbounded geometries (i.e., external flows) and confined internal flows. We demonstrate the success of our Navier-Stokes informed deep learning algorithm by recovering the flow velocity and pressure fields solely from time series data collected on the passive scalar in arbitrary domains.

\section{Problem Setup and Solution Methodology}

Starting with the transport equation \eqref{eq:transC}, we can rewrite it in the following non-dimensional form
\begin{equation}\label{eq:C}
c_t + u c_x + v c_y + w c_z = \text{\footnotesize Pec}^{-1} (c_{xx} + c_{yy} + c_{zz}),
\end{equation}
which governs the evolution of the normalized concentration $c(t,x,y,z)$ of a passive scalar transported by an incompressible Newtonian fluid whose dynamics are described by the Navier-Stokes and continuity equations (also non-dimensional) given below
\begin{equation}\label{eq:NS}
\begin{array}{l}
u_t + u u_x + v u_y + w u_z = - p_x + \text{\footnotesize Re}^{-1} (u_{xx} + u_{yy} + u_{zz}),\\
v_t + u v_x + v v_y + w v_z = - p_y + \text{\footnotesize Re}^{-1} (v_{xx} + v_{yy} + v_{zz}),\\
w_t + u w_x + v w_y + w w_z = - p_z + \text{\footnotesize Re}^{-1} (w_{xx} + w_{yy} + w_{zz}),\\
u_x + v_y + w_z = 0.
\end{array}
\end{equation}
In equations \eqref{eq:C} and \eqref{eq:NS}, $u$, $v$, and $w$ are the $x$, $y$, and $z$ components of the velocity field $\bm{u}=(u,v,w)$, respectively, and $p$ denotes the pressure. A passive scalar is a diffusive field in the fluid flow that has no dynamical effect (such as the effect of temperature in a buoyancy-driven flow) on the fluid motion itself. Smoke and dye are two typical examples of passive scalars. Here, Re represents the Reynolds number, a dimensionless quantity in fluid mechanics, and Pec denotes the P\'{e}clet number, a dimensionless quantity relevant in the study of transport phenomena in flowing fluids. In this work, we assume that the only observables are noisy data $\{t^n,x^n,y^n,z^n,c^n\}_{n=1}^N$ on the concentration $c(t,x,y,z)$ of the passive scalar. Given such data, scattered in space and time, we are interested in inferring the latent (hidden) quantities $u(t,x,y,z)$, $v(t,x,y,z)$, $w(t,x,y,z)$, and $p(t,x,y,z)$. To solve the aforementioned data assimilation problem, we would like to design an algorithm that is agnostic to the geometry as well as the initial and boundary conditions. This is enabling as it will give us the flexibility to work with data acquired in arbitrarily complex domains such as human arteries or brain aneurysms.\\

One question that would naturally arise is whether the information on the passive scalar in the training domain and near its boundaries is sufficient to result in a unique velocity field. The answer is that normally there are no guarantees for unique solutions unless some form of boundary conditions are explicitly imposed on the domain boundaries. However, as shown later for the benchmark problems studied in the current work, an informed selection of the training boundaries in the regions where there are sufficient gradients in the concentration of the passive scalar would eliminate the requirement of imposing velocity and pressure boundary conditions. In addition to the proper design of the training domain and the use of the passive scalar $c$, we can improve the model predictions further by introducing an auxiliary variable $d := 1-c$ (essentially the complement of $c$) that satisfies the transport equation
\begin{equation}\label{eq:D}
d_t + u d_x + v d_y + w d_z = \text{\footnotesize Pec}^{-1} (d_{xx} + d_{yy} + d_{zz}).
\end{equation}
This can be clearly seen by a change of variable from $c$ to $d = 1-c$ and using equation \eqref{eq:C}. The complementary nature of the auxiliary variable $d$ helps the algorithm better detect the geometry and the corresponding boundary conditions. This makes the training algorithm agnostic to the physical geometry of the problem, hence, keeping the implementation and training significantly simpler. In addition, the region of interest becomes very flexible to be chosen with its boundaries no longer required to be the physical boundaries. Another advantage of the use of $d$ is the improvement in the accuracy of the algorithm predictions with no additional cost.\\

\begin{figure}[!t]
\centering
\includegraphics[width=\textwidth]{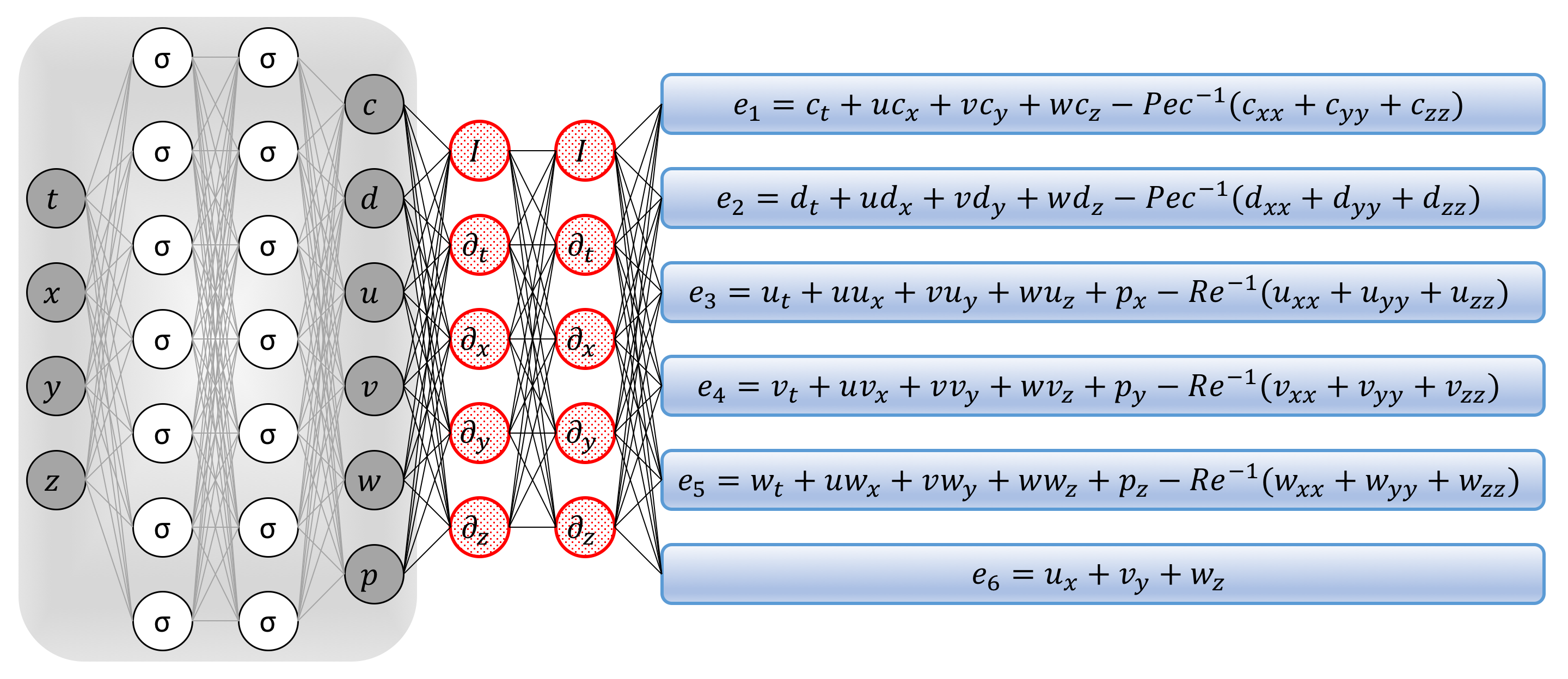}
\caption{\emph{Navier-Stokes informed neural networks:} A plain vanilla densely connected (physics uninformed) neural network, with 10 hidden layers and 50 neurons per hidden layer per output variable (i.e., $6 \times 50 = 300$ neurons per hidden layer), takes the input variables $t, x, y, z$ and outputs $c, d, u, v, w$, and $p$. As for the activation functions, we use $\sigma(x) = \sin(x)$. For illustration purposes only, the network depicted in this figure comprises of 2 hidden layers and 7 neurons per hidden layers. We employ automatic differentiation to obtain the required derivatives to compute the residual (physics informed) networks $e_1$, $e_2$, $e_3$, $e_4$, $e_5$, and $e_6$. If a term does not appear in the blue boxes (e.g., $u_{xy}$ or $u_{tt}$), its coefficient is assumed to be zero. It is worth emphasizing that unless the coefficient in front of a term is non-zero, that term is not going to appear in the actual ``compiled" computational graph and is not going to contribute to the computational cost of a feed forward evaluation of the resulting network. The total loss function is composed of the regression loss of the passive scalar $c$ and its complement $d$ on the training data, and the loss imposed by the differential equations $e_1 - e_6$. Here, $I$ denotes the identity operator and the differential operators $\partial_t, \partial_x, \partial_y$, and $\partial_z$ are computed using automatic differentiation and can be thought of as ``activation operators". Moreover, the gradients of the loss function are backpropogated through the entire network to train the parameters using the Adam optimizer.}\label{fig:BestPhILM}
\end{figure}

Following our earlier work on physics-informed deep learning \cite{raissi2017physics_I, raissi2017physics_II} and \emph{deep hidden physics models} \cite{raissi2018deep}, we approximate the function
\[
(t,x,y,z) \longmapsto (c,d,u,v,w,p)
\]
by a deep neural network and obtain the following \emph{Navier-Stokes informed neural networks} (see figure \ref{fig:BestPhILM}) corresponding to equations \eqref{eq:C}, \eqref{eq:NS}, and \eqref{eq:D}; i.e.,
\begin{equation}\label{eq:NN}
\begin{array}{l}
e_1 := c_t + u c_x + v c_y + w c_z - \text{\footnotesize Pec}^{-1} (c_{xx} + c_{yy} + c_{zz}),\\
e_2 := d_t + u d_x + v d_y + w d_z - \text{\footnotesize Pec}^{-1} (d_{xx} + d_{yy} + d_{zz}),\\
e_3 := u_t + u u_x + v u_y + w u_z + p_x - \text{\footnotesize Re}^{-1} (u_{xx} + u_{yy} + u_{zz}),\\
e_4 := v_t + u v_x + v v_y + w v_z + p_y - \text{\footnotesize Re}^{-1} (v_{xx} + v_{yy} + v_{zz}),\\
e_5 := w_t + u w_x + v w_y + w w_z + p_z - \text{\footnotesize Re}^{-1} (w_{xx} + w_{yy} + w_{zz}),\\
e_6 := u_x + v_y + w_z.
\end{array}
\end{equation}
We acquire the required derivatives to compute the residual networks $e_1$, $e_2$, $e_3$, $e_4$, $e_5$, and $e_6$ by applying the chain rule for differentiating compositions of functions using automatic differentiation \cite{baydin2015automatic}. A schematic representation of the resulting \emph{Navier-Stokes informed neural networks} is given in figure \ref{fig:BestPhILM}. It is worth emphasizing that automatic differentiation is different from, and in several respects superior to, numerical or symbolic differentiation -- two commonly encountered techniques of computing derivatives. In its most basic description \cite{baydin2015automatic}, automatic differentiation relies on the fact that all numerical computations are ultimately compositions of a finite set of elementary operations for which derivatives are known. Combining the derivatives of the constituent operations through the chain rule gives the derivative of the overall composition. This allows accurate evaluation of derivatives at machine precision with ideal asymptotic efficiency and only a small constant factor of overhead. In particular, to compute the required derivatives we rely on Tensorflow \cite{abadi2016tensorflow}, which is a popular and relatively well documented open-source software library for automatic differentiation and deep learning computations. In TensorFlow, before a model is run, its computational graph is defined statically rather than dynamically as for instance in PyTorch \cite{paszke2017automatic}. This is an important feature as it allows us to create and compile the computational graph for the Navier-Stokes informed neural networks \eqref{eq:NN} only once and keep it fixed throughout the training procedure. This leads to significant reduction in the computational cost of the proposed framework.\\

The shared parameters of the neural networks for $c$, $d$, $u$, $v$, $w$, and $p$ can be learned by minimizing the following sum of squared errors loss function
\begin{eqnarray}\label{SSE}
SSE &=& \sum_{n=1}^N |c(t^n,x^n,y^n,z^n) - c^n|^2 \nonumber\\
&+& \sum_{n=1}^N |d(t^n,x^n,y^n,z^n) - d^n|^2 \\
&+& \sum_{i=1}^6 \sum_{n=1}^N |e_i(t^n,x^n,y^n,z^n)|^2.\nonumber
\end{eqnarray}
Here, $d^n = 1 - c^n$ and the first two terms correspond to the training data on the concentration $c(t,x,y,z)$ of the passive scalar while the last term enforces the structure imposed by equations \eqref{eq:C}, \eqref{eq:NS}, and \eqref{eq:D} at a finite set of measurement points whose number and locations are taken to be the same as the training data. However, it should be pointed out that the number and locations of the points on which we enforce the set of partial differential equations could be different from the actual training data. Although not pursued in the current work, this could significantly reduce the required number of training data on the concentration of the passive scalar.

\section{Results}

To generate high-resolution datasets for different benchmark problems studied in the current work, we have employed the spectral/hp-element solver NekTar in which the Navier-Stokes equations \eqref{eq:NS} along with the transport equation \eqref{eq:C} are approximated using high-order semi-orthogonal Jacobi polynomial expansions \cite{karniadakis2013spectral}. The numerical time integration is performed using a third-order stiffly stable scheme until the system reaches its stationary state. In what follows, a small portion of the resulting dataset corresponding to this stationary solution will be used for model training, while the remaining data will be used to validate our predictions. Our algorithm is agnostic to the choice of initial and boundary conditions as well as the geometry. However, we choose to provide this information for every benchmark problem for the sake of completeness and reproducibility of the numerically generated data. Moreover, all data and codes used in this manuscript are publicly available on GitHub at \url{https://github.com/maziarraissi/HFM}.\\

To obtain the results reported in the remainder of this manuscript, we represent each of the functions $c$, $d$, $u$, $v$, $w$, and $p$ by a 10-layer deep neural network with 50 neurons per hidden layer (see figure \ref{fig:BestPhILM}). As for the activation functions, we use $\sin(x)$. In general, the choice of a neural network's architecture (e.g., number of layers/neurons and form of activation functions) is crucial and in many cases still remains an art that relies on one's ability to balance the trade off between \emph{expressivity} and \emph{trainability} of the neural network \citep{raghu2016expressive}. Our empirical findings so far indicate that deeper and wider networks are usually more expressive (i.e., they can capture a larger class of functions) but are often more costly to train (i.e., a feed-forward evaluation of the neural network takes more time and the optimizer requires more iterations to converge). Moreover, the sinusoid (i.e., $\sin(x)$) activation function seems to be numerically more stable than $\tanh(x)$, at least while computing the residual neural networks $e_i$, $i=1,\ldots,6$ (see equation \eqref{eq:NN}). However, these observations should be interpreted as conjectures rather than as firm results\footnote{We encourage the interested reader to check out the codes corresponding to this paper on GitHub at \url{https://github.com/maziarraissi/HFM} and experiment with different choices for the neural networks' architectures.}. In this work, we have tried to choose the neural networks' architectures in a consistent fashion throughout the manuscript. Consequently, there might exist other architectures that could possibly improve some of the results reported in the current work.\\

As for the training procedure, our experience so far indicates that while training deep neural networks, it is often useful to reduce the learning rate as the training progresses. Specifically, the results reported in the following are obtained after $250$, $500$, and $250$ consecutive epochs of the Adam optimizer \cite{kingma2014adam} with learning rates of $10^{-3}$, $10^{-4}$, and $10^{-5}$, respectively. Each epoch corresponds to one pass through the entire dataset. The total number of iterations of the Adam optimizer is therefore given by $1000$ times the number of data divided by the mini-batch size. The mini-batch size we used is $10000$ and the number of data points are clearly specified in the following on a case by case basis. Every $10$ iterations of the optimizer takes around $1.9$ and $3.8$ seconds, respectively, for two and three dimensional problems on a single NVIDIA Titan X GPU card.

\subsection{External flows}

\begin{figure}[!t]
\centering
\includegraphics[width=0.8\textwidth]{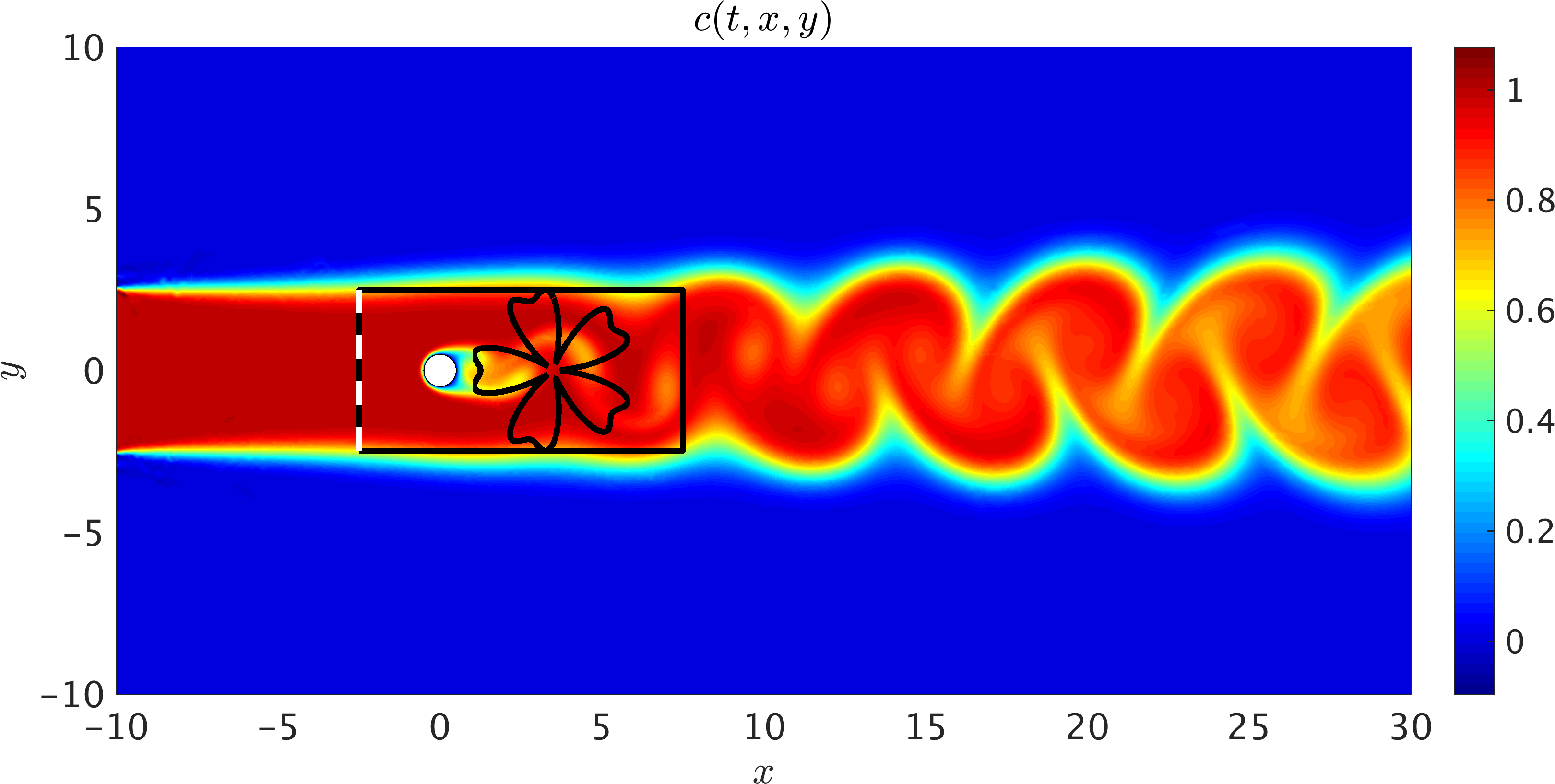}
\caption{\emph{2D Flow past a circular cylinder:} 2D sketch of the simulation domain in which data on the concentration field is generated. We have assumed a uniform free stream velocity profile at the left boundary, a zero pressure outflow condition imposed at the right boundary located $30D$ $(D=1)$ downstream of the cylinder, and periodicity for the top and bottom boundaries of the $[-10,30]\times[-10,10]$ domain. The passive scalar is injected at the inlet from [-2.5, 2.5]. A completely arbitrary training domain in the shape of a flower is depicted in the wake of the cylinder. No information on the velocity is given for this training domain. Another training domain is also shown by a rectangle that includes the cylinder. Information on the velocity is only given on the left boundary of the rectangular domain shown by the dashed line.}\label{fig:cylinder_2D_flower}
\end{figure}

As a first example, we consider the prototypical problem of a two dimensional flow past a circular cylinder, known to exhibit rich dynamic behavior and transitions for different regimes of the Reynolds number $\text{Re} = U_{\infty} D/\nu$. Assuming a non-dimensional free stream velocity $U_{\infty} = 1$, cylinder diameter $D = 1$, and kinematic viscosity $ \nu = 0.01$, the system exhibits a periodic steady state behavior characterized by an asymmetrical vortex shedding pattern in the wake of the cylinder, known as the K\'{a}rm\'{a}n vortex street \cite{batchelor2000introduction}. Importantly, the passive scalar is injected at the inlet within the interval [-2.5, 2.5] using a step function (see figure \ref{fig:cylinder_2D_flower}). The choice of this boundary condition only depends on the region of interest in which velocity and pressure fields are inferred using the concentration field. Furthermore, a constant value of diffusivity for the passive scalar is assumed to be given by $\kappa = 0.01$ resulting in $\text{Pec} = U_{\infty} \kappa/\nu = 100$. Here, we use zero-slip and zero-concentration boundary conditions on the cylinder wall. Whereas the P\'eclet number is chosen to be equal to the Reynolds number, there is no restriction on its value as shown for the internal flow benchmark problem. Physically, for the flow of gases such as air with smoke as a passive scalar $\text{Pec} \approx \text{Re}$. This is not, however, the case for liquid flows with dye as the passive scalar since diffusivity of dyes are typically smaller than most fluids, which leads to higher P\'eclet numbers.\\

\begin{figure}[!t]
\includegraphics[width=\textwidth]{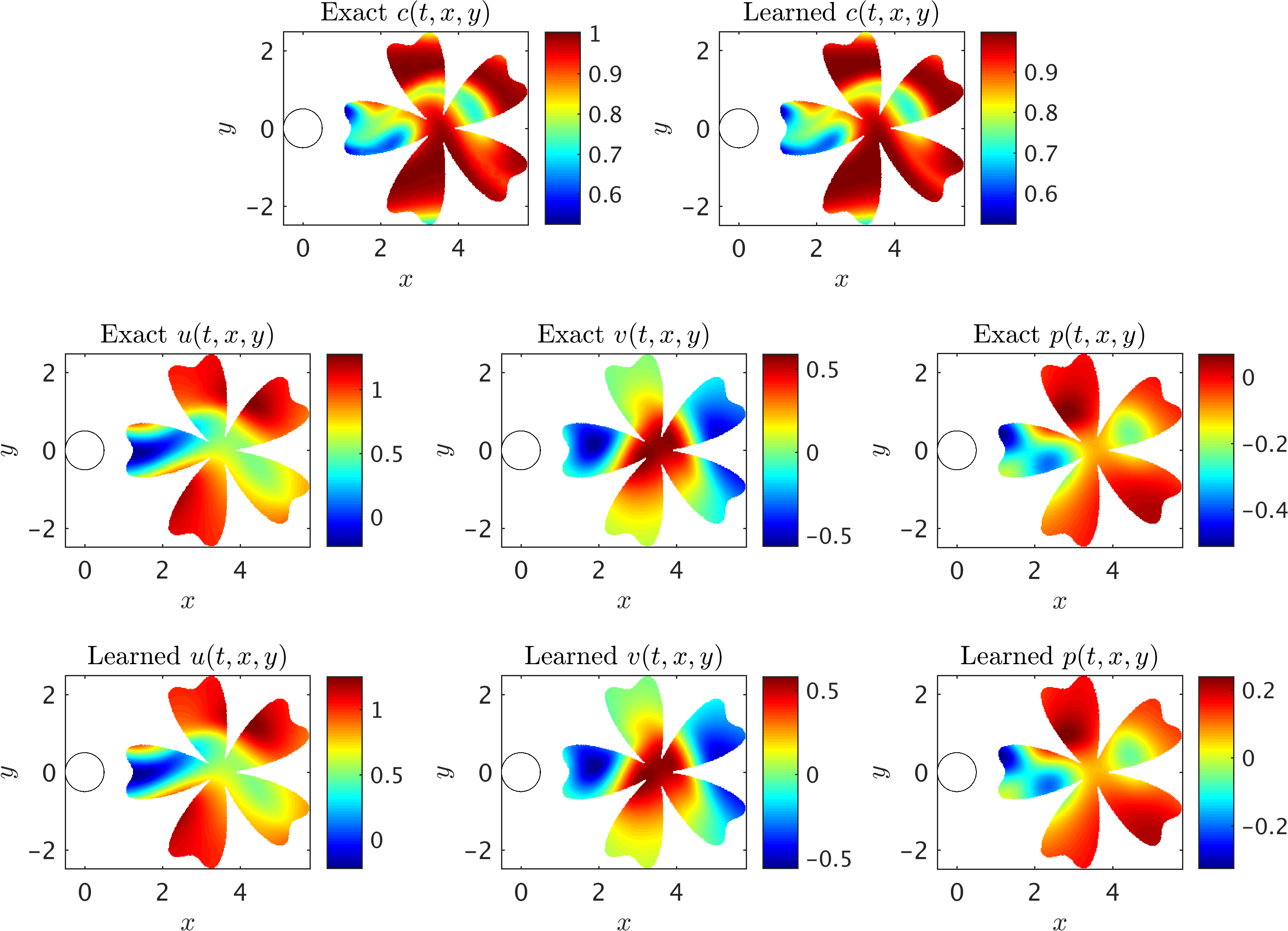}
\caption{\emph{2D Flow past a circular cylinder:} A representative snapshot of the input data on the concentration of the passive scalar is shown in the top left panel of this figure, where on the right the same concentration field is reconstructed based on the predictions of our algorithm. The algorithm is capable of accurately reconstructing the velocity $u, v$ and the pressure $p$ fields shown in the third row. The exact velocity and pressure fields at the same point in time are plotted for comparison in the second row. Note that the pressure is off by a constant since this is an incompressible flow.}\label{fig:cylinder_2D_flower_results}
\end{figure}

\begin{figure}[!t]
\includegraphics[width=\textwidth]{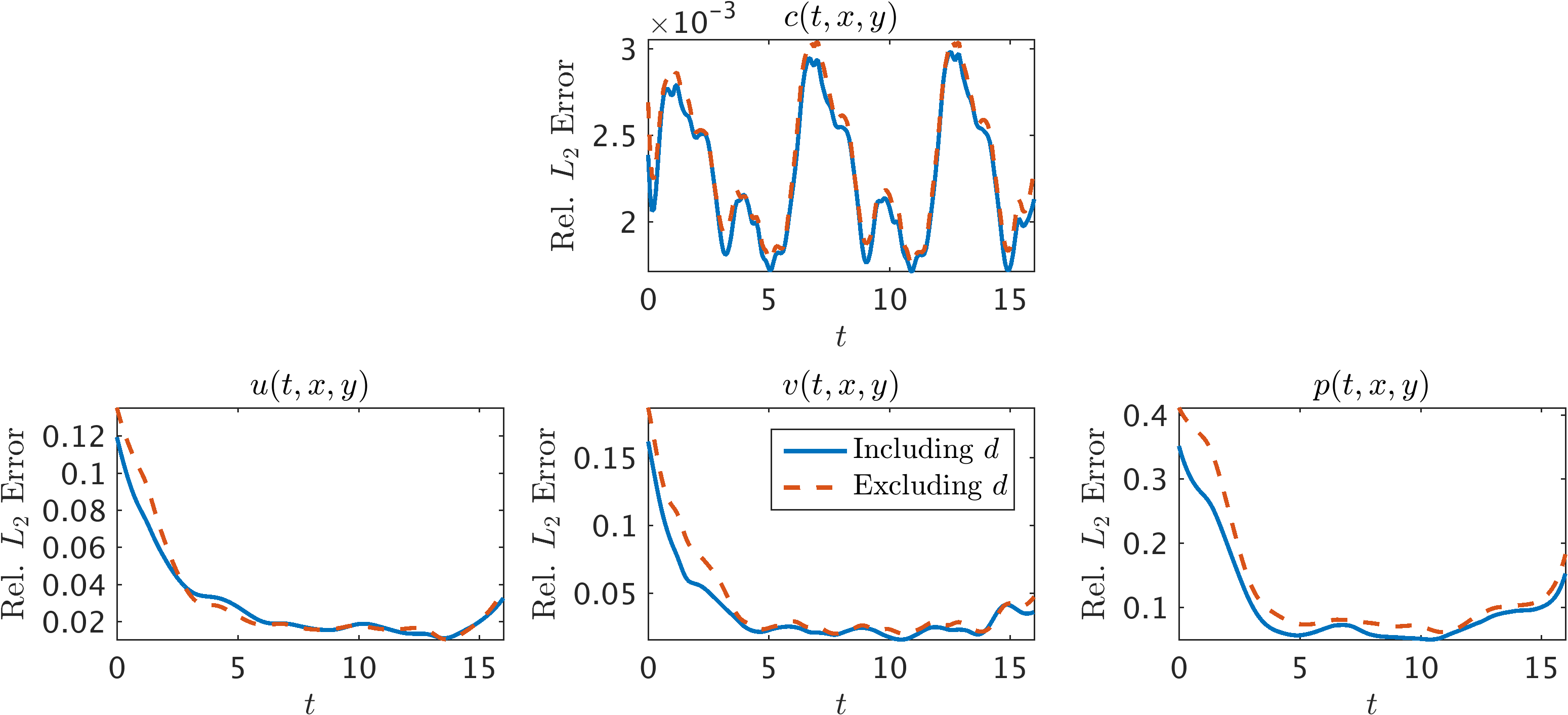}
\caption{\emph{2D Flow past a circular cylinder:} Relative $L_2$ errors between predictions of the model and the corresponding exact concentration, velocity, and pressure fields for the arbitrary training domain downstream of the cylinder. 31 million data points, scattered in space and time, are used both to regress the concentration field and enforce the corresponding partial differential equations. Furthermore, the training is performed with the auxiliary variable $d$ (blue line) and without $d$ (red dashed line) for comparison.}\label{fig:Cylinder2D_flower_results_errors}
\end{figure}

A representative snapshot of the input data on the concentration field is shown in figure \ref{fig:cylinder_2D_flower}. As illustrated in this figure, the shape and extent of the boundaries of the training domains that we choose for our analysis could be \emph{arbitrary} and may vary by problem. However, it should be pointed out that there are two important factors that need to be considered when choosing the training domain. First, the concentration field of the passive scalar must be present within the training domain, such that its information can be used to infer other flow variables. Second, to avoid the need for specifying appropriate boundary conditions for velocities, there must exist enough concentration gradient normal to the boundaries (i.e., $\partial c/\partial n \neq 0$) in order for our method to be able to infer a unique solution for the velocity field. As shown in figure \ref{fig:cylinder_2D_flower_results}, excellent agreement can be achieved between the predictions of our algorithm and the exact data within a completely arbitrary training domain downstream of the cylinder, while the inclusion of auxiliary variable $d=1-c$ can improve the accuracy of predictions further (see figure \ref{fig:Cylinder2D_flower_results_errors}). Here, no input information other than the concentration field for the passive scalar $c$ is passed to the algorithm.\\

\begin{figure}[!t]
\includegraphics[width=\textwidth]{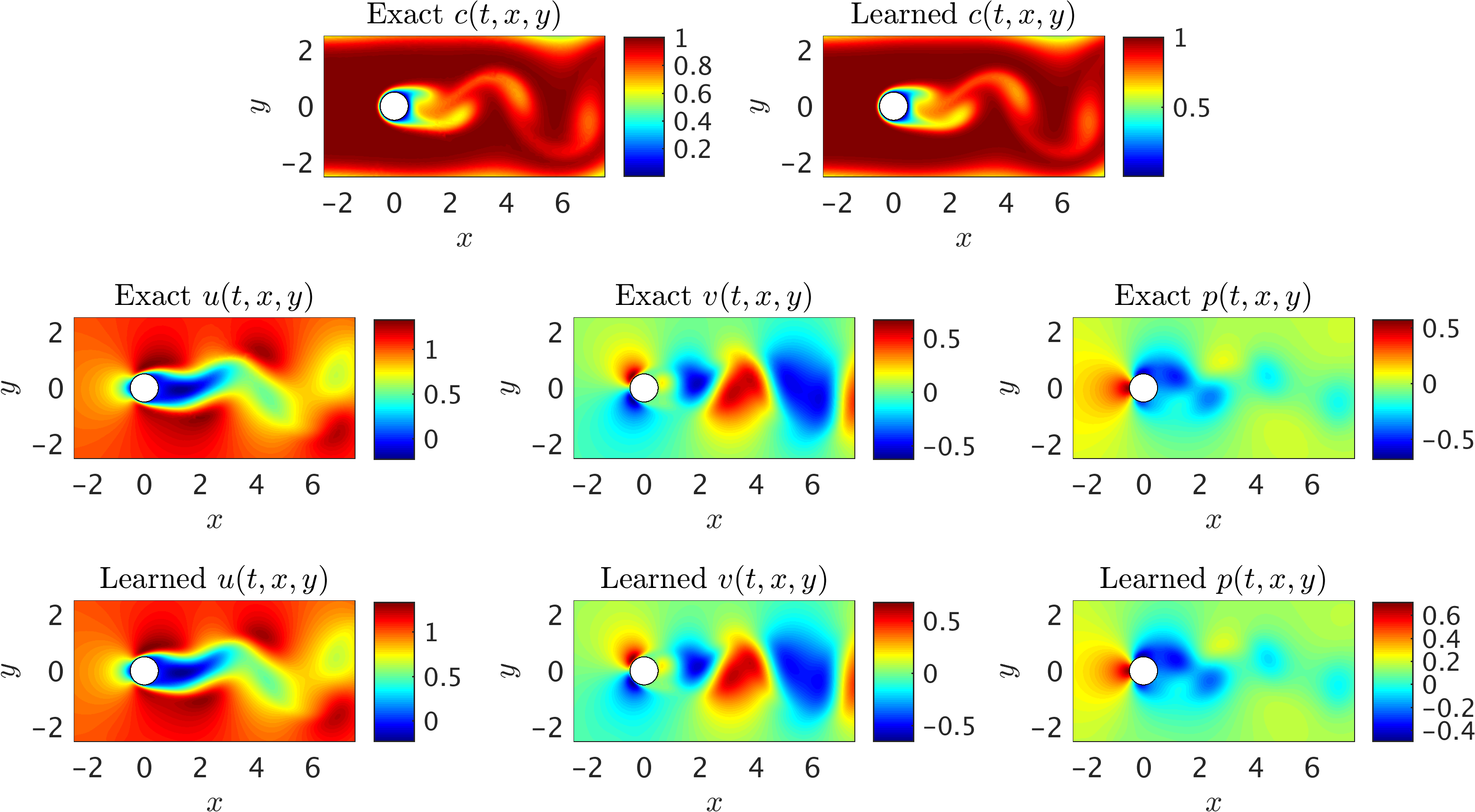}
\caption{\emph{2D Flow past a circular cylinder:} A representative snapshot of the input data on the concentration of the passive scalar is shown in the top left panel of this figure, where on the right the same concentration field is reconstructed based on the predictions of our algorithm. The algorithm is capable of accurately reconstructing the velocity $u, v$ and the pressure $p$ fields shown in the third row. The exact velocity and pressure fields at the same point in time are plotted for comparison in the second row. Note that the pressure is off by a constant since this is an incompressible flow.}\label{fig:cylinder_2D_Pec100_included_results}
\end{figure}

Next, we choose the region of interest to contain the cylinder (see the rectangular domain in figure \ref{fig:cylinder_2D_flower}) so that the fluid forces acting on the cylinder can be inferred. As shown in figure \ref{fig:cylinder_2D_Pec100_included_results}, the algorithm is capable of accurately reconstructing the velocity and the pressure fields without having access to sufficient observations of these fields themselves. Note that we use a very small training domain that cannot be used in classical computational fluid dynamics to obtain accurately the Navier-Stokes solutions. It should be emphasized that other than the velocity on the left boundary, no other boundary conditions are given to the algorithm. We need to impose a Dirichlet boundary condition for the velocity at the left boundary simply because the observations of the concentration are not providing sufficient information (i.e., gradients) in front of the cylinder. More notably, there is no need to impose the no-slip condition on the cylinder wall. The presence of the normal concentration gradient naturally allows our algorithm to infer the zero velocity condition on the walls. In regards to the predicted pressure field, we note that due to the nature of the Navier-Stokes equations for incompressible flows, the pressure field is only identifiable up to a constant.\\

\begin{figure}[!t]
\centering
\includegraphics[width=\textwidth]{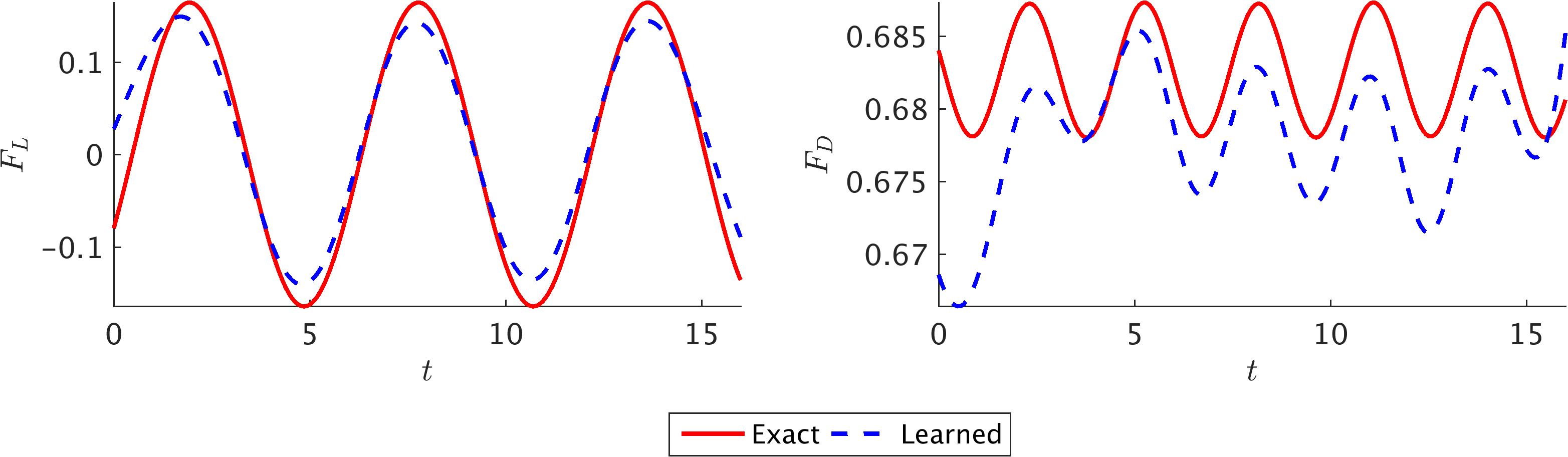}
\caption{\emph{Non-dimensional lift and drag forces on 2D cylinder:} Comparison between the exact (spectral element solution) and predicted lift (left) and drag (right) forces per unit length of the cylinder.}\label{fig:lift_drag}
\end{figure}

\begin{figure}[!t]
\includegraphics[width=\textwidth]{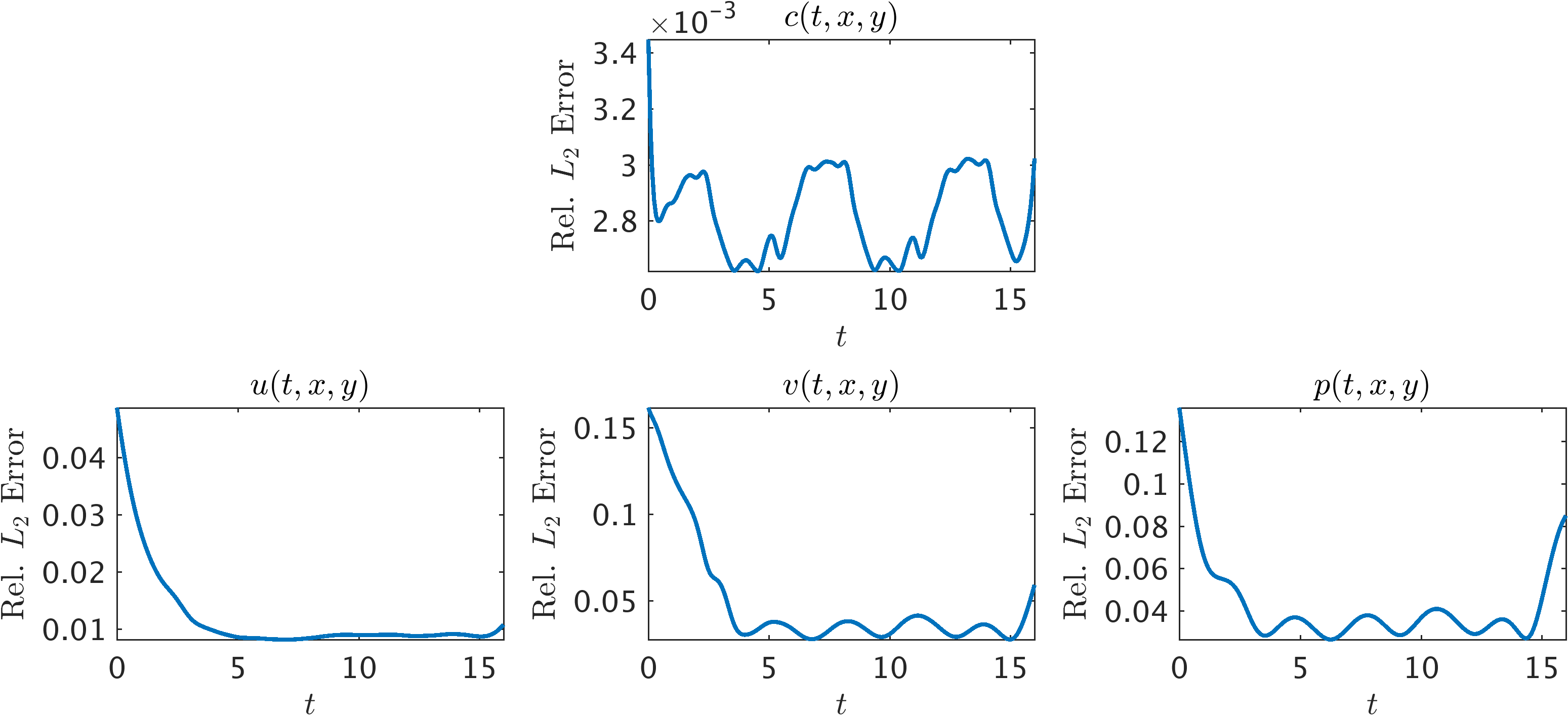}
\caption{\emph{2D Flow past a circular cylinder:} Relative $L_2$ errors between predictions of the model and the corresponding exact concentration, velocity, and pressure fields for the rectangular training domain that contains the cylinder. 2.8 million data points scattered within the training domain and in time are used for both regressing the concentration field and enforcing the corresponding partial differential equations.}\label{fig:Cylinder2D_results_errors}
\end{figure}

The learned pressure and velocity fields can be used to obtain the lift and drag forces exerted on the cylinder by the fluid as shown in figure \ref{fig:lift_drag}. The fluid forces on the cylinder are functions of the pressure and velocity gradients. Consequently, having trained the neural networks, we can use
\begin{eqnarray*}
F_L = \oint \left[-p n_y + 2 \text{\footnotesize Re}^{-1} v_y n_y + \text{\footnotesize Re}^{-1} \left(u_y + v_x\right)n_x\right]ds,\\
F_D = \oint \left[-p n_x + 2 \text{\footnotesize Re}^{-1} u_x n_x + \text{\footnotesize Re}^{-1} \left(u_y + v_x\right)n_y\right]ds,
\end{eqnarray*}
to obtain the lift and drag forces, respectively. Here, $\bm{n} = (n_x,n_y)$ is the outward normal on the cylinder and $ds$ is the arc length on the surface of the cylinder. We use the trapezoidal rule to approximately compute these integrals. Interestingly, the predicted lift and drag forces are in excellent agreement with the exact ones, both in terms of the frequency of oscillations and the amplitude. The resulting error is within 1\% of the test data. Some discrepancy, however, can be observed for the few initial and final time instants, which can be attributed to the lack of data. This can be further clarified by the relative $\mathcal{L}_2$-norm of error results shown in figure \ref{fig:Cylinder2D_results_errors}. Lack of training data on $c$ for $t < 0$ and $t > 16$ leads to weaker neural network predictions for the initial and final time instants. Thus, one should take this into consideration when inference is required within a certain time interval. Note that to compute lift and drag forces, the gradient of the velocity has to be computed on the wall, which can be done analytically using the parametrized surrogate velocity field (i.e., the neural networks). Although no numerical differentiation is needed to compute the gradients, the integration of forces on the surface of the cylinder is approximated by a summation.\\

In addition to the velocity and pressure fields, it is possible to discover other unknown parameters of the flow field such as the Reynolds and P\'{e}clet numbers. Although these parameters were prescribed in the 2D flow past the cylinder example, we have tested a case in which both parameters are free to be learned by the algorithm. The results are given in table \ref{tab:cylinder_2D_RePec}, which shows very good agreement with the exact values. From the practical standpoint, the passive scalar diffusivity and the fluid viscosity (hence, their ratio i.e., the Prandtl number $\text{Pr} \equiv \nu/\kappa$) may be known in advance. Therefore, discovering the Reynolds number would be sufficient whereas the P\'{e}clet number can be computed by Pec = Re Pr.\\

\begin{table}
\caption{\emph{2D flow past a circular cylinder:} Learned Reynolds and P\'{e}clet numbers considered as free parameters of the model.}\label{tab:cylinder_2D_RePec}
\begin{center}
\footnotesize
\begin{tabular}{ |c|c|c|c| } 
 \hline
  & Exact & Learned & Rel. Error \\
  \hline
  \hline
 Pec & 100 & 92.39 & 7.60\% \\ 
 \hline
 Re & 100 & 92.47 & 7.52\% \\ 
 \hline
\end{tabular}
\end{center}
\end{table}

\begin{figure}[!t]
\centering
\includegraphics[width=0.7\textwidth]{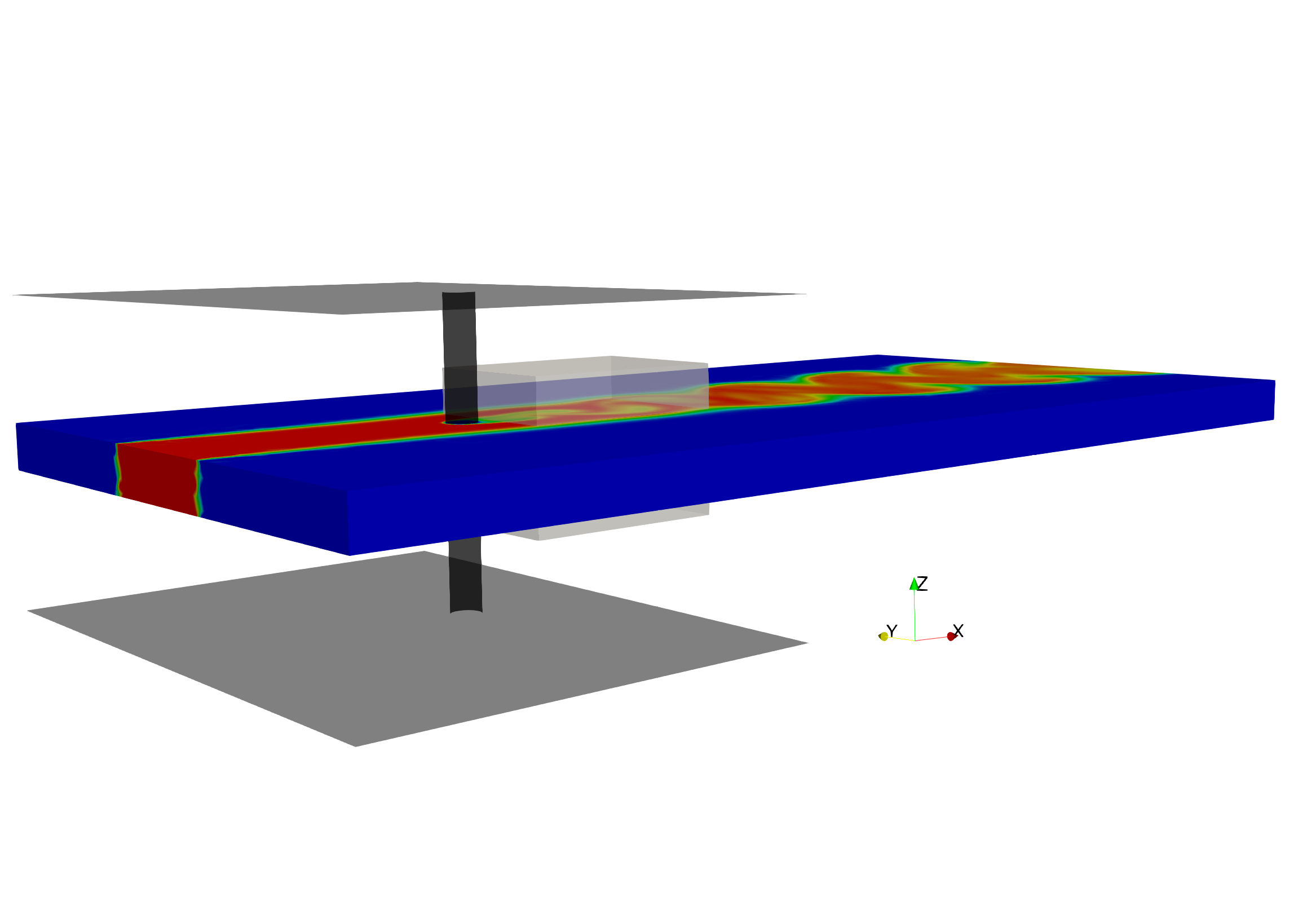}
\caption{\emph{3D flow past a circular cylinder:} 3D sketch for the prototypical problem of flow past a circular cylinder of diameter $D = 1$. Two parallel planes are located $10D$ apart along the $z-$axis. Flow is bounded between the plates for $10D$ along $x-$axis. Periodic boundary conditions are prescribed parallel to the $xz$ plane at $y = -10$ and $10$, and parallel to the $xy$ plane at $z = 0$ and $10$ starting from the location where the walls end. All the physical boundaries are shown in black, whereas the gray box located behind the cylinder shows the domain of interest where model training is performed. Zero-Neumann boundary conditions are imposed for velocities and concentration along with zero pressure at the outflow located $30D$ downstream of the cylinder. A uniform $U_{\infty} = 1$ is imposed at the inlet located $8D$ upstream of the cylinder. Passive scaler is injected at the inlet through a finite region $[-2.5, 2.5]\times[0, 10]$. Zero-velocity and concentration boundary conditions are imposed on each physical boundary. Note that no information on the surfaces of the training (gray box) domain is given.}\label{fig:cylinder3d}
\end{figure}

The previous benchmark example was a two-dimensional (2D) flow, where we could safely neglect the $z$-coordinate and $w$-component of the velocity from the input and output variables, respectively. For the flow past a cylinder, if we simply increase the Reynolds number beyond a threshold value of $\approx 185$ \cite{ma2002low}, the spanwise velocity $w$ becomes more prominent due to the effect of the so-called ``vortex stretching'' \cite{batchelor2000introduction,karniadakis1992three}. To test the capability of the proposed algorithm in inferring three-dimensional (3D) flow fields, we design another prototype problem of the 3D flow past a finite-size circular cylinder confined between two parallel plates as shown in figure \ref{fig:cylinder3d}, where similar to the previous example, we set $\text{Re} = \text{Pec} = 100$. Downstream of the cylinder, the flow exits to an open region, which causes strong 3D effects in the wake of the cylinder. Hence, we set the training domain in the wake of cylinder.\\

\begin{figure}[!t]
\centering
\includegraphics[width=0.8\textwidth]{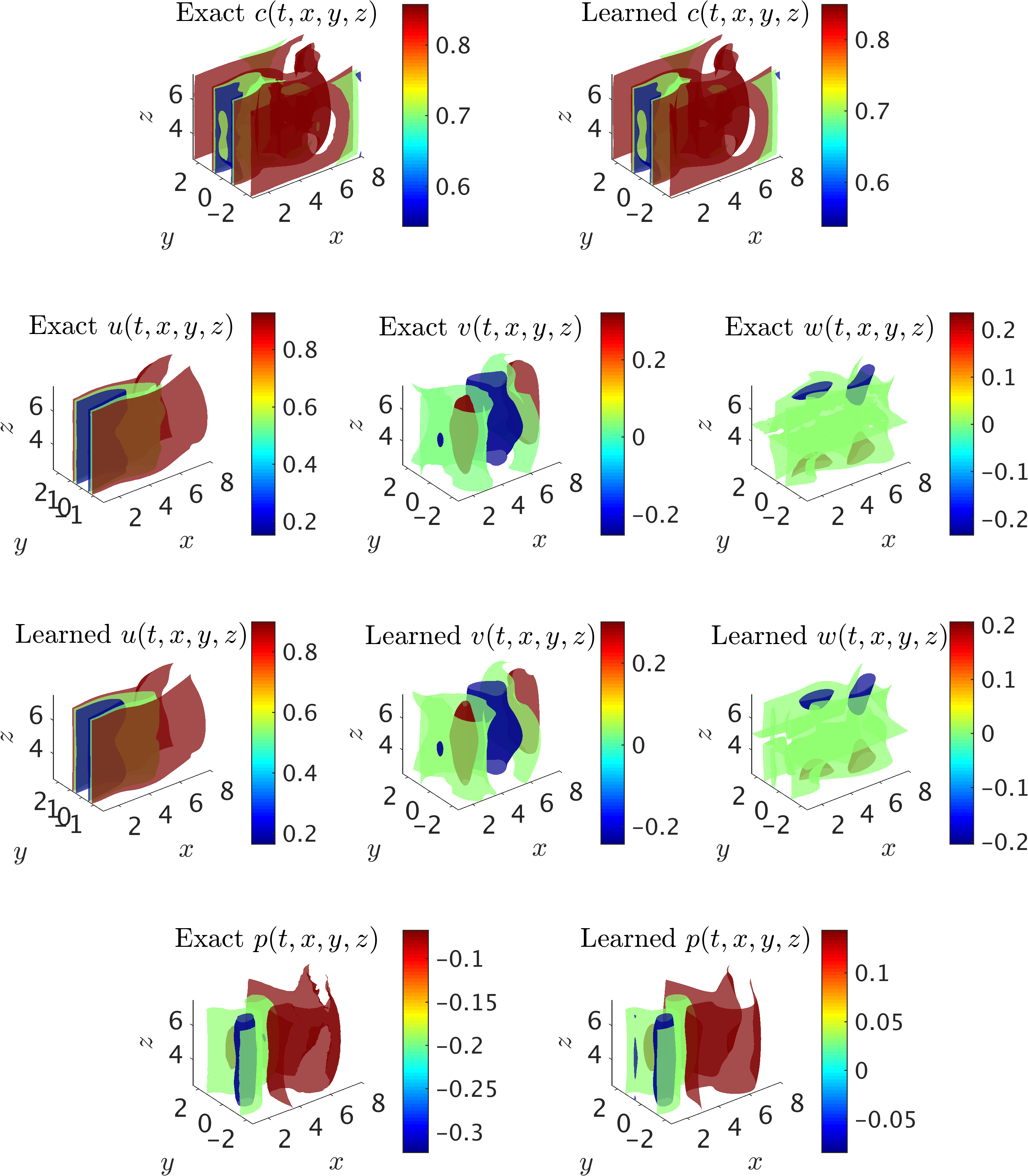}
\caption{\emph{3D flow past a circular cylinder:} The iso-surfaces of exact data and the predictions of our algorithm on the concentration of the passive scalar are shown in top row for a representative time instant within the selected training domain. Using the information on the concentration only, the velocity fields $u, v, w$ are inferred, are shown in the third row, and are compared with the exact data in the second row. In addition, the exact and predicted pressure $p$ fields are plotted in the last row. Note the range of contour levels are set equal between the exact and predicted iso-surfaces of concentration fields and velocity components for better comparison.}\label{fig:cylinder_3D_Pec100_results}
\end{figure}

\begin{figure}[!t]
\centering
\includegraphics[width=0.9\textwidth]{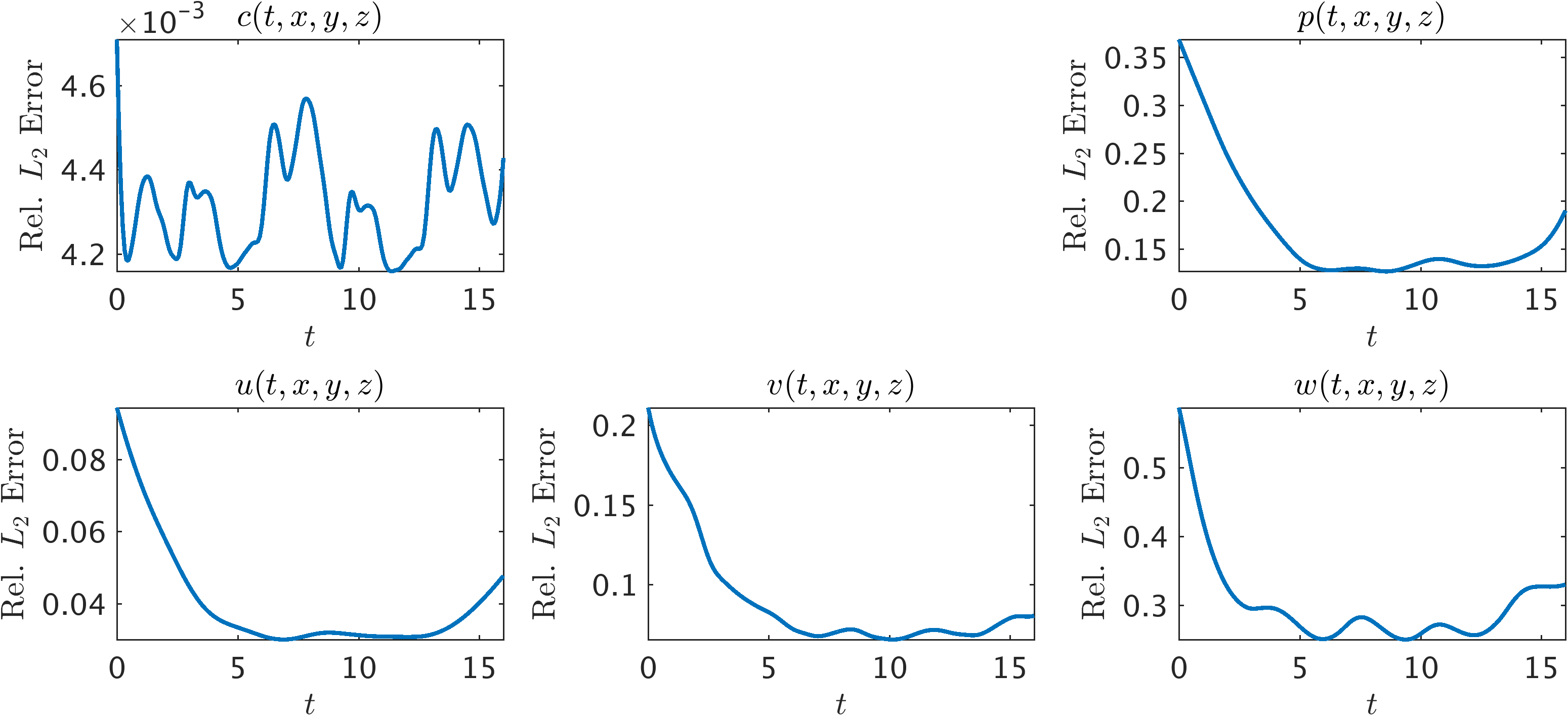}
\caption{\emph{3D flow past a circular cylinder:} Relative $L_2$ errors between predictions of the model and the corresponding exact concentration, velocity, and pressure fields. 37 million data points scattered within the training box and in time are used for both regressing the concentration field and enforcing the corresponding partial differential equations.}\label{fig:Cylinder3D_errors}
\end{figure}

A representative snapshot of the input data on the concentration field in the wake of the cylinder is plotted in the top left panel of figure \ref{fig:cylinder_3D_Pec100_results}. The iso-surfaces for all of the fields are also plotted for comparison between the exact data and the predictions of our algorithm. The algorithm is capable of accurately reconstructing the velocity and the pressure fields without having access to any observations of these fields. In particular, no information on the velocity is given on the boundaries of the domain of interest in the training phase. Qualitative comparison between the predictions of our algorithm and the exact data shows good agreement for 3D flows as well, however, the relative $\mathcal{L}_2$-norm errors for velocity and pressure fields are slightly higher than the values for the 2D flow predictions as shown in figure \ref{fig:Cylinder3D_errors}. Similar to 2D results, the neural network training lack sufficient data for the initial and final time instants that leads to larger errors in the predictions.

\subsection{Internal flows}
\begin{figure}[ht]
\centering
\includegraphics[width=\textwidth]{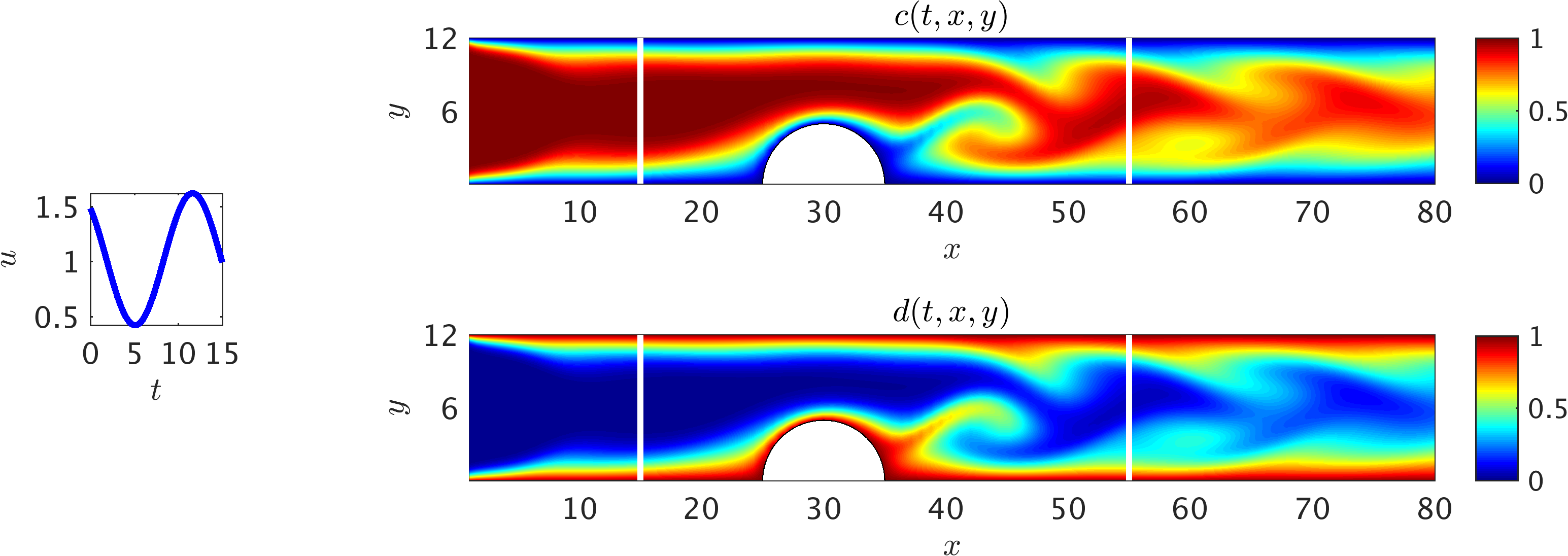}
\caption{\emph{2D channel flow over an obstacle:} Contours of the concentration field within a 2D channel at a representative time instant. A pulsatile velocity profile $u(t)$ ($v=0$) is imposed at the inlet (shown as the inset on the left), and a uniform concentration $c = 1$ for the passive scalar is injected to the channel. At the outlet, a zero-Neumann boundary condition for the velocity and concentration is considered, while the pressure is set to zero. Furthermore, the velocity and concentration on the walls are set to zero. The flow $\text{Re} = \bar{U} H/\nu = 60$ is calculated based on the mean inflow velocity $\bar{U}=1$ and channel hight $H=12$, whereas we choose a smaller diffusion constant for the passive scalar leading to higher $\text{Pec} = \bar{U} H/\kappa = 180$. The training domain (white rectangle) is considered to contain the obstacle with the upper and lower boundaries being the physical wall boundaries. Also shown are contours of the auxiliary variable $d$, which signifies the important role of this variable in helping the algorithm detect the boundaries of the training domain.}\label{fig:Stenosis2D_data}
\end{figure}

We now turn our attention to an important class of flows in confined geometries also known as ``internal flows". While measuring average flow velocity and pressure in ducts, pipes and even blood vessels is now a common practice, quantifying the spatial fields, specifically the shear stresses on the boundaries, is not a trivial task. The first benchmark problem considered here is the transient flow over an obstacle in a 2D channel as shown in figure \ref{fig:Stenosis2D_data}. To make the flow unsteady, we impose a velocity waveform at the inlet of the channel. Passive scalar value on the boundaries is set to zero, where the boundaries are assumed to be impenetrable. It is clear that the auxiliary variable $d := 1-c$ is equal to one on the channel boundaries. In other words, while the passive scalar concentration $c$ is convected downstream from the inflow, the complement scalar $d$ is ``virtually" infused from the boundaries to the flow stream as shown in figure \ref{fig:Stenosis2D_data}. The presence of the obstacle breaks the symmetry in the flow for which analytical solutions do not exist. Thus, to estimate the velocity field, direct measurements or forward numerical simulations are required.\\

\begin{figure}[!t]
\centering
\includegraphics[width=\textwidth]{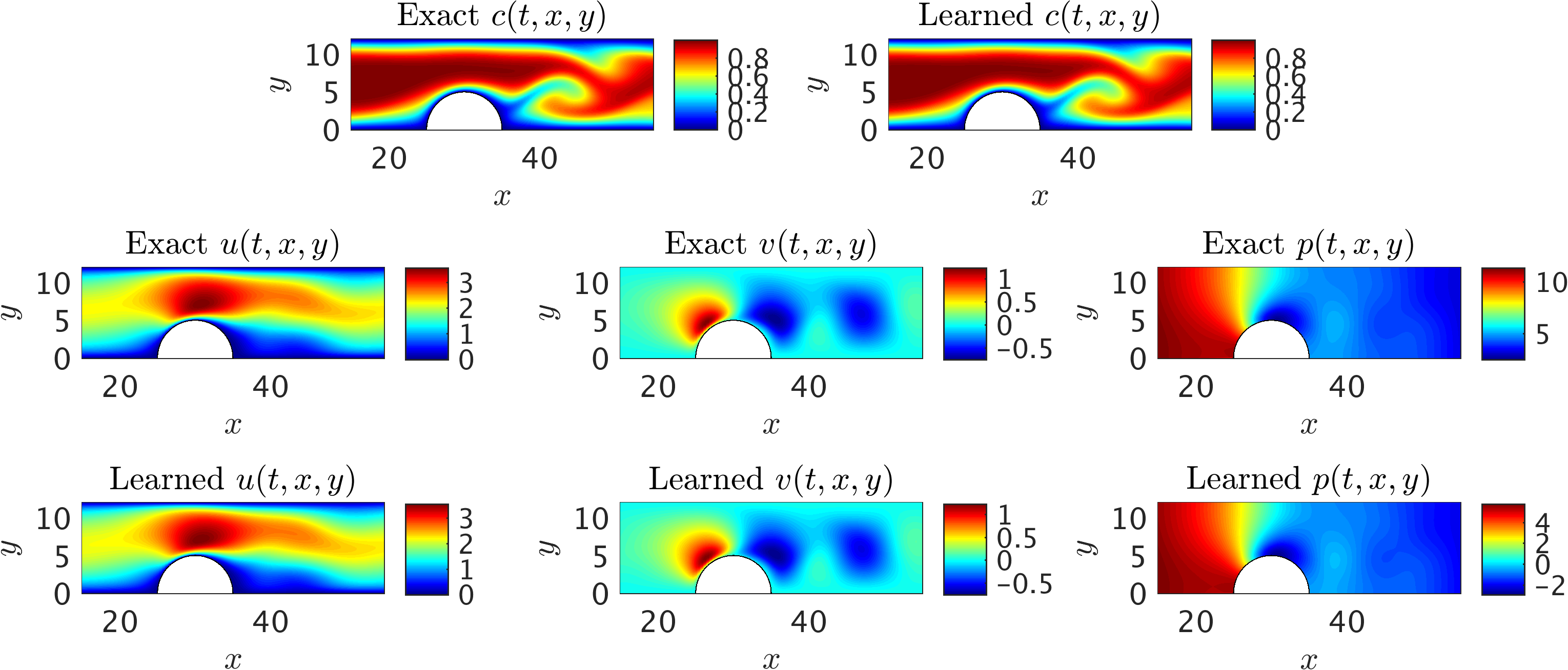}
\caption{\emph{2D channel flow over an obstacle:} A representative snapshot of the input data on the concentration field within the training domain is plotted in the top left panel alongside the prediction of our algorithm. The algorithm is capable of accurately reconstructing the velocity and the pressure fields without having access to any observations of these fields (shown in the second and third rows). Furthermore, no boundary conditions are specified on the boundaries of the training domain including the physical wall boundaries.}\label{fig:stenosis_2D_included_results}
\end{figure}

\begin{figure}[!t]
\centering
\includegraphics[width=\textwidth]{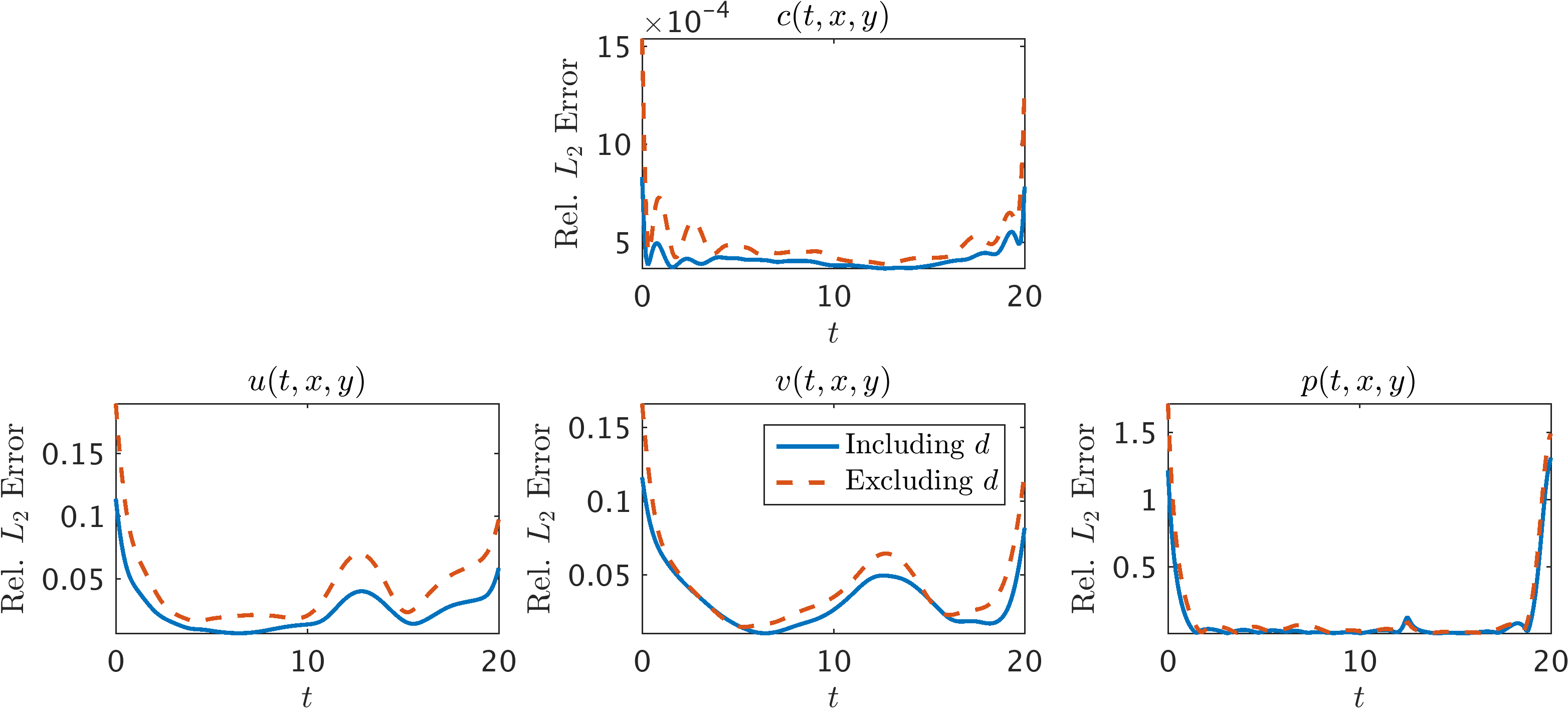}
\caption{\emph{2D channel flow over an obstacle:} Relative $L_2$ errors between predictions of the model and the corresponding exact concentration, velocity, and pressure fields. 10 million data points scattered within the training domain and in time are used for both regressing the concentration field and enforcing the corresponding partial differential equations. Furthermore, the training is performed with the auxiliary variable $d$ (blue line) and without $d$ (red dashed line) for comparison.}\label{fig:Stenosis2D_results_errors}
\end{figure}

A representative snapshot of the input data on the concentration field along with its neural network approximation are depicted in the top panel of figure \ref{fig:stenosis_2D_included_results}. As shown in figure \ref{fig:stenosis_2D_included_results}, predictions of the algorithm for velocity and pressure fields are in excellent agreement with the exact data. Note that no information for velocity is given on the boundaries of the channel or the obstacle. It should be mentioned that while the concentration gradients normal to the physical boundaries are sufficient to infer the flow variables without the knowledge of the boundary conditions (i.e., no-slip velocity on the wall), the \emph{free} information from the auxiliary variable $d$ in addition to $c$ helps the algorithm improve the accuracy of predictions (shown in figure \ref{fig:Stenosis2D_results_errors}) at practically no additional cost.\\

One of the real-world examples of the above benchmark problem is encountered in cardiovascular fluid mechanics, where one or multiple coronary arteries are partially blocked by atherosclerosis plaques formed by the lipid accumulation \cite{hansson2005inflammation}. An accurate prognosis of these ``stenotic'' vessels is an extremely important step in the treatment planning and decision-making process for patients with cardiovascular disease \cite{budoff2008diagnostic}. Direct measurements of pressure in the vessel is invasive and bears high risk, whereas computational fluid modeling of blood in the coronaries requires correct reconstruction of the geometry as well as the knowledge of all the inflow/outflow boundary conditions. The proposed inverse approach to infer the velocity and pressure fields with the use of a passive scalar, however, offers a promising alternative to the conventional methods. Here, the passive scalar could be the bolus dye that is typically injected to the blood stream for the purpose of blood flow monitoring and medical imaging.\\

\begin{figure}[!t]
\centering
\includegraphics[width=\textwidth]{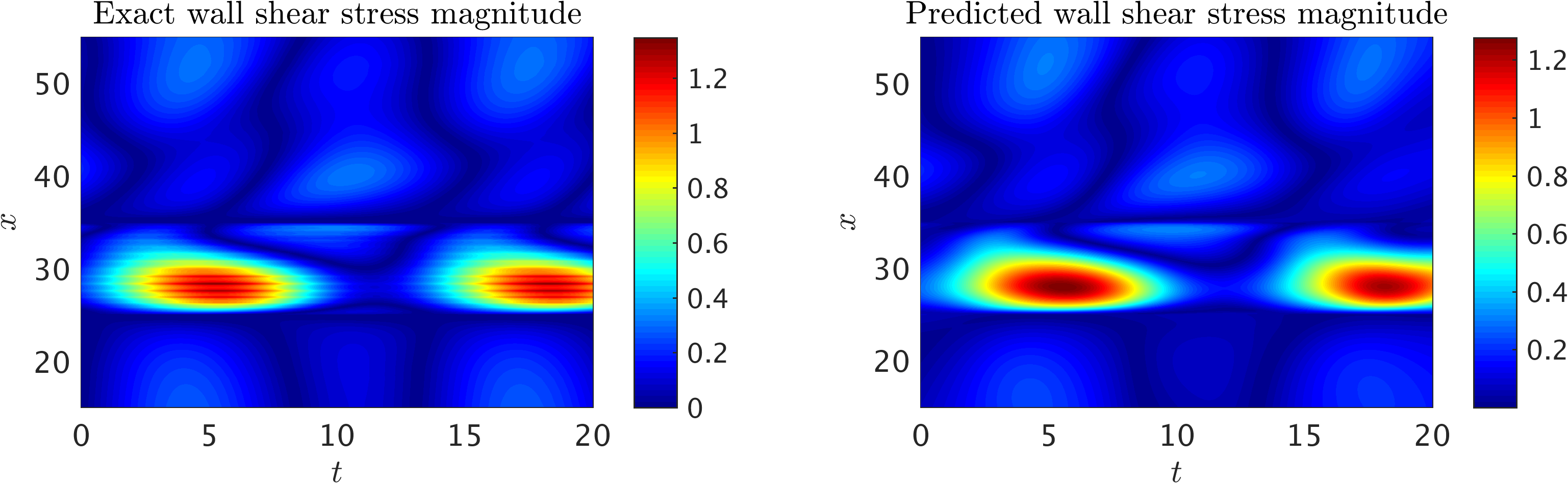}
\caption{\emph{2D channel flow over an obstacle:} Wall shear stress magnitude, $\text{WSS}(x,t)$, computed on the lower wall ($15 \leq x \leq 55$) in the training domain of figure \ref{fig:Stenosis2D_data} as a function of time ($0 \leq t \leq 20$). Strong wall shear stress values are observed on the surface of the cylinder ($25 \leq x \leq 35$). The plot also shows the periodicity in the flow field.}\label{fig:Stenosis2D_results_shears}
\end{figure}

Using the predicted velocity fields, we are able to compute shear stresses everywhere in the training domain. Of particular interest are wall shear stresses, which can be computed using the following equations in 2D; 
\[
\begin{array}{l}
\tau_{x} = 2 \text{\footnotesize Re} \left[u_x n_x + \frac{1}{2}(v_x + u_y)n_y\right],\\
\tau_{y} = 2 \text{\footnotesize Re} \left[\frac{1}{2}(u_y + v_x) n_x + v_y n_y\right].
\end{array}
\]
Here, $\bm{n} = (n_x,n_y)$ is the outward normal on the boundary of the domain. Note that to compute wall shear stresses $\bm{\tau} = (\tau_{x}, \tau_{y})$, the gradient of the velocity is required, which can be computed using automatic differentiation. Wall shear stresses are important quantities of interest in many biological processes e.g., in the pathogenesis and progression of vascular diseases that cause aortic aneurysms. Here, we have estimated the temporal wall shear stress magnitudes, $\text{WSS}(x,t) = \sqrt[]{\tau_{x}^2 + \tau_{y}^2}$, acting on the lower wall using the predictions of neural networks for the 2D channel flow over the obstacle, and plotted them against the results from the spectral/hp element solver in figure \ref{fig:Stenosis2D_results_shears}. The results show excellent agreement between the predictions and numerical estimations. Note that the numerical estimations suffer from a slight aliasing effect (noisy oscillations in the $x$ direction) for which dealiasing is required to retrieve a smooth distribution of wall shear stresses. Interestingly, the neural networks prediction is smooth as the velocity gradients of a parameterized velocity field are taken analytically. Furthermore, similar to the previous benchmark problems, slight discrepancies can be observed close to the initial and final times ($t \sim 0,\ 20$) due to the lack of training data. This is further explained in figure \ref{fig:Stenosis2D_results_errors}, where relative $\mathcal{L}_2$-norm of errors in the velocity fields show a peak close to the initial and final time instants out of the entire prediction time interval. Also shown in the figure is the enhancement in the model prediction accuracy when the auxiliary variable $d$ is used as opposed to when $d$ is not included during the training phase.\\

Similar to the external flow problem past the cylinder, the algorithm is able to learn the Reynolds and P\'{e}clet numbers as free parameters. The predicted values of these two parameters are given in table \ref{tab:Stenosis2D_RePec}, which shows excellent agreement with the exact values. Note that unlike the flow past the cylinder, the physical Re and Pec numbers are not the same. Remarkably, the algorithm is capable of inferring the underlying physics without knowledge of initial and boundary conditions and physical boundaries, where the information is provided only on the transported passive scalar.\\

\begin{table}[!t]
\caption{\emph{2D channel flow over an obstacle:} Learned Reynolds and P\'{e}clet numbers considered as free parameters of the model.}\label{tab:Stenosis2D_RePec}
\begin{center}
\footnotesize
\begin{tabular}{ |c|c|c|c| } 
 \hline
  & Exact & Learned & Rel. Error \\
  \hline
  \hline
 Pec & 180 & 178.95 & 0.58\% \\ 
 \hline
 Re & 60 & 59.92 & 0.12\% \\ 
 \hline
\end{tabular}
\end{center}
\end{table}

\begin{figure}[!t]
\centering
\includegraphics[width=\textwidth]{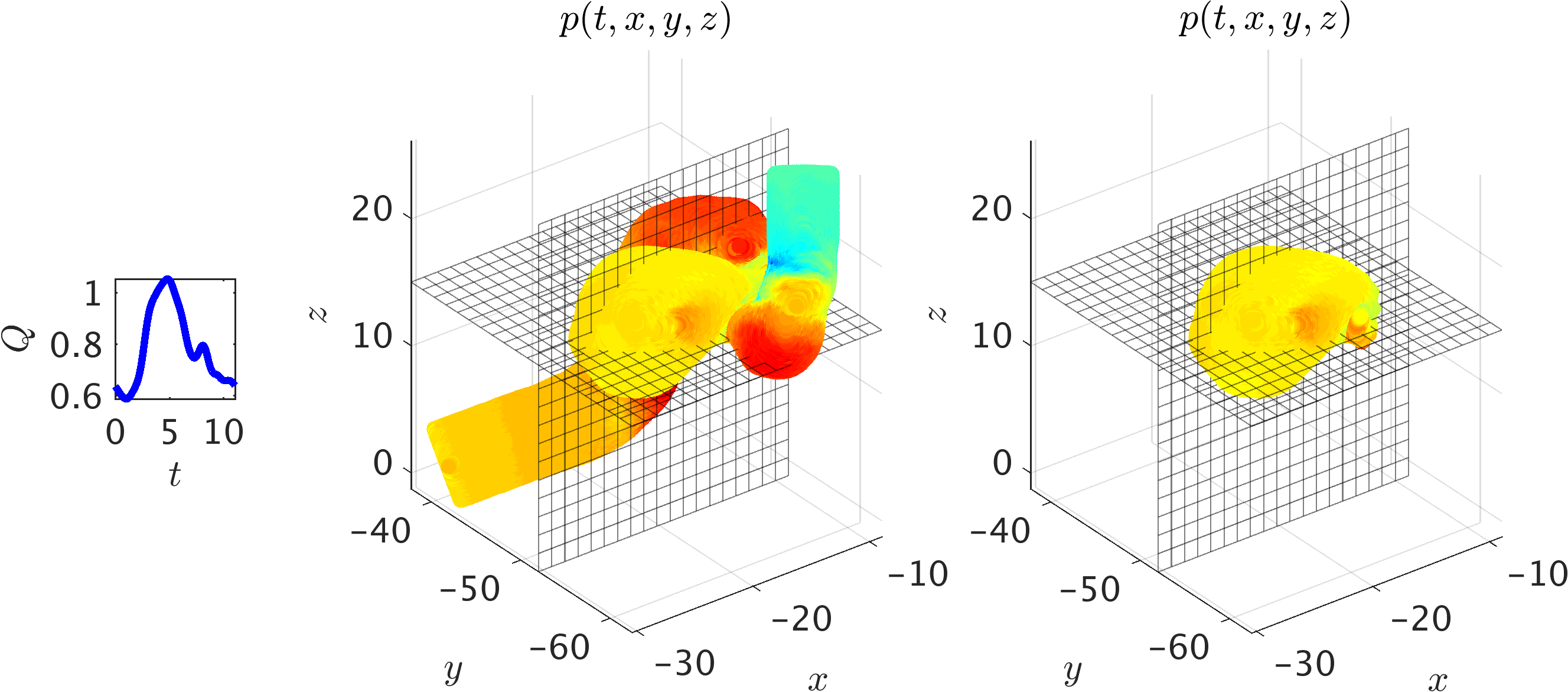}
\caption{\emph{A 3D intracranial aneurysm:} The middle panel shows the simulation domain and pressure field at a time instant, whereas on the right the training domain containing only the ICA sac is shown. Two perpendicular planes have been used to interpolate the exact data and the predicted ones for plotting 2D contours in figure \ref{fig:ICA_results}. A physiologic flow waveform $Q(t)$ (shown in the inset figure) is prescribed at the inlet along with the uniform concentration for the passive scalar. At the outlet, zero-Neumann boundary conditions are imposed for velocities and concentration, whereas a ``Windkessel'' type boundary condition is used for the pressure to represent the truncated geometry downstream \cite{grinberg2008outflow}. The Reynolds and P\'{e}clet numbers are estimated based on the mean velocity and lumen diameter at the inlet. Using the kinematic viscosity of blood and assuming a diffusivity constant $\kappa$ for the passive scalar equal to the viscosity, we obtain $\text{Re} = \text{Pec} = 98.2$.}\label{fig:ICA_data}
\end{figure}

To further illustrate the implications of the \emph{Navier-Stokes informed neural networks} in addressing real-world problems, we consider 3D physiologic blood flow in a realistic intracranial aneurysm (ICA) shown in figure \ref{fig:ICA_data}. The aneurysm is located in the cavernous segment of the right internal carotid artery at the level of the eye and beneath the brain \cite{baek2012convergence}. Exact concentration fields are generated numerically using realistic boundary conditions, which is a physiologic flow waveform at the inlet along with a uniform concentration for the passive scalar. The strength of the proposed algorithm is its ability to stay agnostic with respect to the geometry as well as initial and boundary conditions. Hence, it is possible to focus only on the regions where velocity and pressure fields are needed, which will significantly reduce the size of data and the cost of training.\\

\begin{figure}[!t]
\centering
\includegraphics[width=\textwidth]{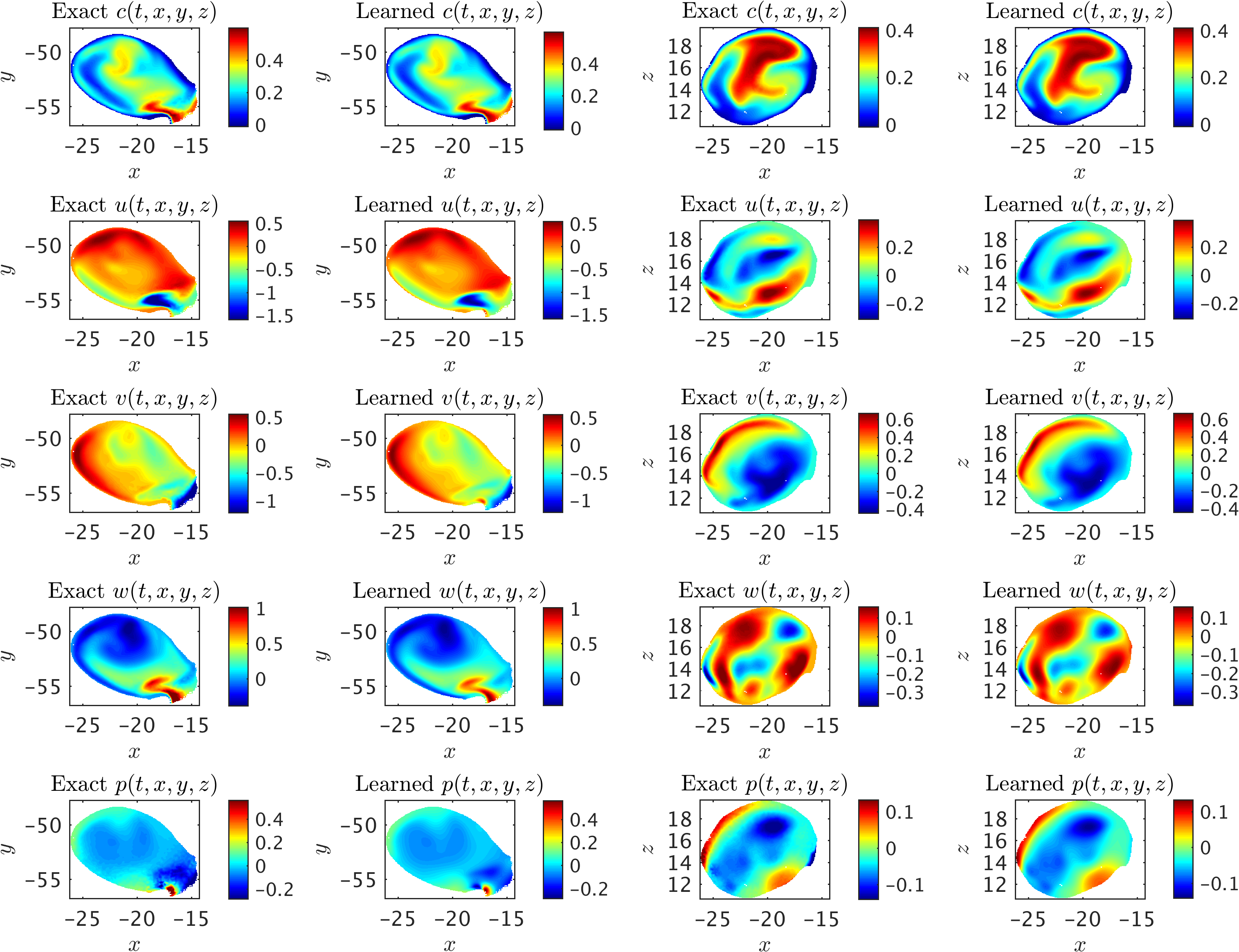}
\caption{\emph{A 3D intracranial aneurysm:} Contours of the exact fields and model predictions are plotted on two perpendicular planes (shown in figure \ref{fig:ICA_data}) for concentration $c$, velocity $u, v, w$, and pressure $p$ fields in each row. The first two columns show the results interpolated on a plane perpendicular to the $z$-axis, and the next two columns are plotted for a plane perpendicular to the $y$-axis. Note that the range of contour levels are set equal for all fields for better comparison.}\label{fig:ICA_results}
\end{figure}

\begin{figure}[!t]
\centering
\includegraphics[width=\textwidth]{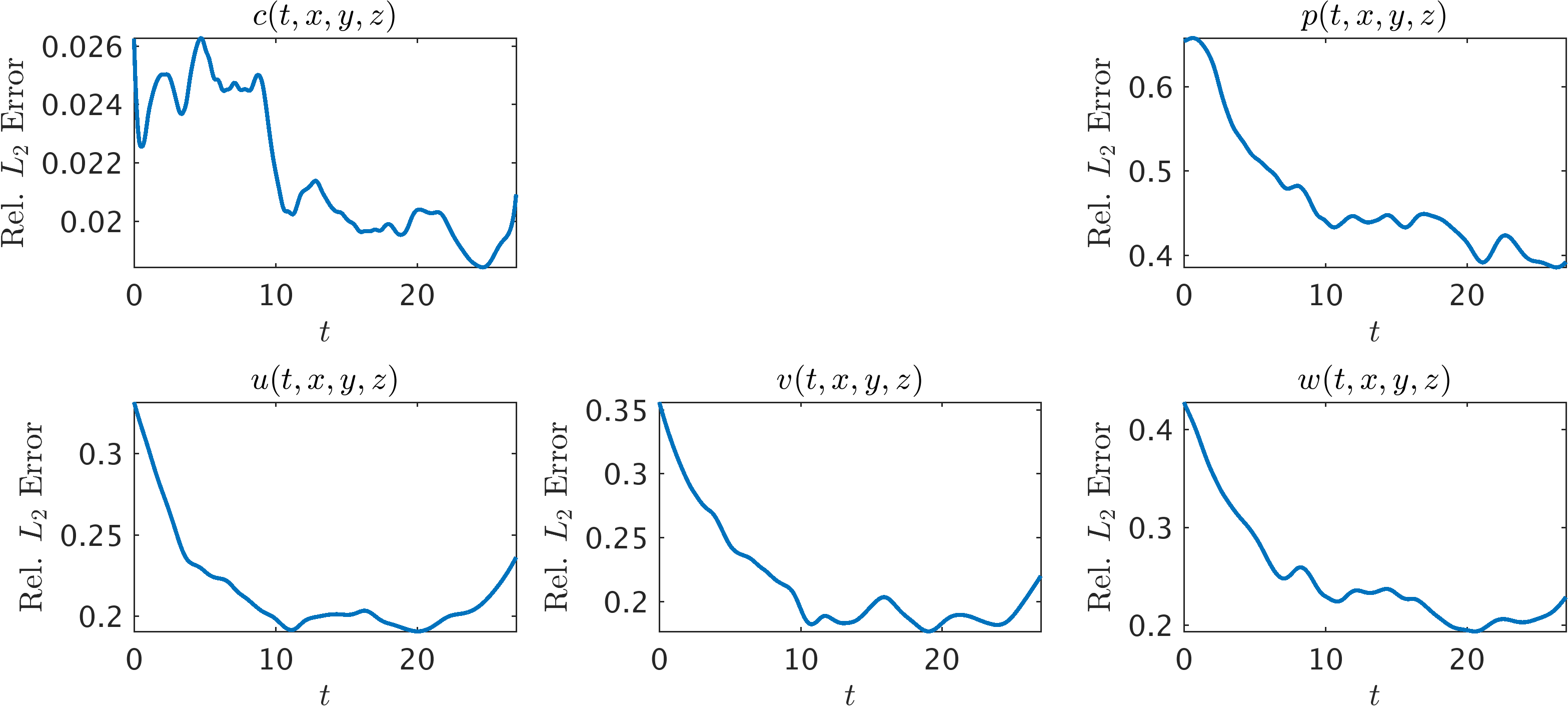}
\caption{\emph{A 3D intracranial aneurysm:} Relative $L_2$ errors between predictions of the model and the corresponding exact concentration, velocity, and pressure fields. 29 million data points scattered within the aneurysm sac and in time are used for both regressing the concentration field and enforcing the corresponding partial differential equations.}\label{fig:ICA_errors}
\end{figure}

We first crop the aneurysm sac out from the rest of geometry, and then use only the scalar data within the ICA sac (right panel of figure \ref{fig:ICA_data}) for training where no information is used for the boundary conditions. The exact and predicted concentration, velocity and pressure fields within the ICA sac at a sample time instant are then interpolated on two separate planes perpendicular to $y$- and $z$-axis, which are shown in figure \ref{fig:ICA_results}. We observe excellent agreement between the exact and predicted fields given the complexity of the flow field. Furthermore, the relative $\mathcal{L}_2$-norm of errors plotted in figure \ref{fig:ICA_errors} show a significant drop in the prediction errors away from the initial and final time instants, which justifies an accurate analysis of model predictions in the time interval $ 10 \leq t \leq 25$.

\section{Discussion and Concluding Remarks}

The algorithm developed here is agnostic to the geometry, initial, and boundary conditions, which makes it highly flexible in choosing the domain of interest for data acquisition as well as subsequent training and predictions. Moreover, the current methodology allows us to construct computationally efficient and fully differentiable surrogates for velocity and pressure fields that can be further used to estimate other quantities of interest such as shear stresses and vorticity fields. The predictions presented here are among those cases where a pure machine learning algorithm or a mere scientific computing approach simply cannot reproduce. A pure machine learning strategy has no sense of the physics of the problem to begin with, and a mere scientific computing approach relies heavily on careful specification of the geometry as well as initial and boundary conditions.\\

Assuming the geometry to be known, to arrive at similar results as the ones presented in the current work, one needs to solve significantly more expensive optimization problems, using conventional computational methods (e.g., finite differences, finite elements, finite volumes, spectral methods, and etc.). The corresponding optimization problems involve some form of ``parametrized" initial and boundary conditions, appropriate loss functions, and multiple runs of the conventional computational solvers. In this setting, one could easily end up with very high-dimensional optimization problems that require either backpropagating through the computational solvers \cite{chen2018neural} or ``Bayesian" optimization techniques \cite{shahriari2016taking} for the surrogate models (e.g., Gaussian processes). If the geometry is further assumed to be unknown (as is the case in this work), then its parametrization requires grid regeneration, which makes the approach almost impractical.\\

It must be mentioned that we are avoiding the regimes where the Navier-Stokes equations become chaotic and turbulent (e.g., as the Reynolds number increases). In fact, it should not be difficult for a plain vanilla neural network to approximate the types of complicated functions that naturally appear in turbulence. However, as we compute the derivatives required in equation \eqref{eq:NN}, minimizing the loss function \eqref{SSE} might become a challenge \cite{raissi2018deep}, where the optimizer may fail to converge to the right values for the parameters of the neural networks. It might be the case that the resulting optimization problem inherits the complicated nature of the turbulent Navier-Stokes equations. Hence, inference of turbulent velocity and pressure fields could be considered in future extensions of this line of research.\\

The emphasis of this study was to demonstrate, through several prototypical and realistic examples, the ability of the current algorithm to infer the hidden states of the system from partial knowledge of some relevant quantities by leveraging the known underlying dynamics of the system. We have verified that with the use of governing physical laws, our algorithm is able to make accurate predictions for complex 2D/3D flows. One possible limitation of the current study is the use of synthetically generated data on the passive scalar, which are relatively noiseless and clean compared to the realistic measurements. Whereas 3D reconstruction of a passive scalar from a stack of 2D projections is possible, it is not a common practice in the clinical and industrial settings due to the technological complexities as well as computational and processing costs. Thus, extracting information from 2D images directly by taking into account the angle of projections seems to be a more realistic approach. Furthermore, different imaging modalities (e.g., magnetic resonance vs. computed tomography angiography of bolus dye in coronary arteries) have different spatial/temporal resolutions. To address the lack of resolution in time, a viable strategy is to use a high-order time-stepping scheme such as implicit Runge-Kutta for resolving the equations in time using very few snapshots \cite{raissi2017physics_I,raissi2017physics_II}. These are certainly important factors that have to be taken into account moving forward using the proposed algorithm for real-world fluid mechanics applications, and will be addressed carefully in future implementations.\\

In this work we have been operating under the assumption of Newtonian and incompressible fluid flow governed by the Navier-Stokes equations. However, the proposed algorithm can also be used when the underlying physics is non-Newtonian, compressible, or partially known. This in fact is one of the advantages of the algorithm in which other unknown parameters such as the Reynolds and P\'eclet numbers can be inferred in addition to the velocity and pressure fields. When the fluid is non-Newtonian (e.g., blood flow in small vessels), one can encode the momentum equations, where the ``constitutive" law for the fluid's stress-strain relationship is unknown, into a physics-informed deep learning algorithm. Having data on the velocity (or even the passive scalar), it should be possible to learn the constitutive law as well \cite{zhao2018active}.\\

As demonstrated in this work, a direct implication of the current method is quantifying the hemodynamics in the vasculature. This could potentially have significant impact on the clinical diagnosis (especially with the non-invasive methods) of vascular diseases associated with important pathologies such as heart attack and stroke. Blood shear stresses acting on the vascular wall are crucial in the prognosis of a vascular disease and their quantification is significantly important clinically \cite{shaaban2000wall,zarins1983carotid}. Using the proposed method, it is possible to estimate the wall shear stresses at no extra cost. This will simplify the complexities of the state-of-the-art methods in which extracting the exact boundaries of the vessels from the clinical images is required \cite{boussel2009phase}. In addition to the shear stresses estimation that is of interest, quantification of blood distribution and oxygenation in cerebral arteries using modalities such as perfusion CT \cite{wintermark2002prognostic} and functional magnetic resonance imaging \cite{calamante1999measuring} has been practiced and is crucial in the assessment of brain activity for the majority of neurodegenerative diseases such as Alzheimer’s disease \cite{greicius2004default} and in cognitive neuroscience \cite{aron2007triangulating}. Furthermore, recent advances in imaging such as the functional photoacoustic microscopy allows 3D blood oxygenation imaging with ``capillary-level" resolution \cite{yao2015high}. Encoding the one-dimensional hemodynamics and oxygen transport equations into physics-informed neural networks, and using the concentration profiles from different imaging modalities will lead to quantifying brain hemodynamics at various length scales.\\

In conclusion, we have introduced an effective algorithm based on the recently developed \emph{physics-informed neural networks} framework that is capable of encoding the underlying physical laws that govern a given dataset. Specifically, we have applied the algorithm to an important class of physical laws governing fluid motions. Our inverse approach in this article is unique as we have designed a methodology to infer velocity and pressure fields merely from the knowledge of the time-evolution of a passive scaler.

\section*{Acknowledgements}
This work received support by the DARPA EQUiPS grant N66001-15-2-4055, the AFOSR grant FA9550-17-1-0013, the NIH grant U01HL116323, and the NSF grant DMS-1736088. Generation of data was performed on XSEDE resources supported by award No.~TG-DMS140007. Moreover, all data and codes used in this manuscript will be publicly available on GitHub at \url{https://github.com/maziarraiss/HFM}.





\bibliographystyle{model1-num-names}
\bibliography{sample.bib}







\end{document}